\documentclass{aa}
\usepackage{graphicx}
\usepackage{times}
\usepackage{astron}
\usepackage{psfig}

\def\hide#1{}

\begin{document}

\thesaurus{03.13.2; 03.13.6; 03.20.4; 04.19.1}

\title{Object Classification in Astronomical Multi-Color Surveys}

\author{C. Wolf \and K. Meisenheimer \and H.-J. R\"oser}
\institute{
 Max-Planck-Institut f\"ur Astronomie, K\"onigstuhl 17,
   D-69117 Heidelberg, Germany}

\date{Received 5 Apr 2000 / Accepted}
\maketitle

%\psdraft

\begin{abstract}
We present a photometric method for identifying stars, galaxies and quasars in multi-color surveys, which uses a library of $\ga 65000$ color templates for comparison with observed objects. The method aims for extracting the information content of object colors in a statistically correct way, and performs a classification as well as a redshift estimation for galaxies and quasars in a unified approach based on the same probability density functions. For the redshift estimation, we employ an advanced version of the Minimum Error Variance estimator which determines the redshift error from the redshift dependent probability density function itself. 

The method was originally developed for the Calar Alto Deep Imaging Survey (CADIS), but is now used in a wide variety of survey projects. We checked its performance by spectroscopy of CADIS objects, where the method provides high reliability (6 errors among 151 objects with $R<24$), especially for the quasar selection, and redshifts accurate within $\sigma_z \approx 0.03$ for galaxies and $\sigma_z \approx 0.1$ for quasars.

For an optimization of future survey efforts, a few model surveys are compared, which are designed to use the same total amount of telescope time but different sets of broad-band and medium-band filters. Their performance is investigated by Monte-Carlo simulations as well as by analytic evaluation in terms of classification and redshift estimation. If photon noise were the only error source, broad-band surveys and medium-band surveys should perform equal, as long as they provide the same spectral coverage. In practice, medium-band surveys show superior performance due to their higher tolerance for calibration errors and cosmic variance.

Finally, we discuss the relevance of color calibration and derive important conclusions for the issues of library design and choice of filters. The calibration accuracy poses strong constraints on an accurate classification, which are most critical for surveys with few, broad and deeply exposed filters, but less severe for surveys with many, narrow and less deep filters. %
\end{abstract}

\section{Introduction}

Sky surveys are designed to provide statistical samples of astronomical objects, aiming for spatial overview, completeness and homogeneous datasets. Mostly they serve as a database for rather general conclusions about abundant objects, but another attractive role is allowing to search for rare and unusual objects. For both purposes, it is very useful to predict rather precisely the appearance of the different known types of objects. The object types can then be discriminated successfully, and allow to extract the information content from the survey. Also, unusual objects can be found as inconsistent with all known sorts of objects, but they might as well hide among the bulk of normal objects mimicking their appearance.

In this picture, we of course want a survey to perform as reliable and as accurate as possible in measuring object characteristics like class, redshift or physical parameters. Since surveys aim typically for large samples upon which future detailed work is based, their results are often not extremely reliable and accurate for a given single object. But for a statistical analysis of large samples, we can usually do without perfect accuracy in the measurement of features and we can also accept occasional misclassifications. 

In astronomical surveys pointing off the galactic plane, obvious classes to start out with could basically be {\it stars}, {\it galaxies}, {\it quasars} and {\it strange} objects. These can be further differentiated into subclasses, based on physical characteristics derived from their morphology or spectral energy distribution (SED). Therefore, morphology and color or prominent spectral features are the typical observational criteria applied to survey data for classifying the objects contained. 

Presently, surveys concentrate mostly on either imaging or spectroscopy. While spectroscopic surveys deliver a potentially high spectral resolution, they have expensive requirements for telescope time. Imaging multi-color surveys can expose a number of filters consecutively, and deliver {\it morphological} information and crude {\it spectral} information for all objects contained in the field of view.

Since the subject of this paper is the {\it spectral} information in multi-color surveys, we want to mention {\it morphological} information only briefly: The morphology is only of limited use for classifying objects into stars, galaxies and quasars: Objects observed as clearly extended are certainly not single stars, but the smaller ones could either be galaxies, low-luminosity quasars, or chance projections of more than one object. Objects consistent with point-sources can be stars, compact galaxies or quasars. Also, the morphological differentiation depends on the seeing conditions and typically reaches not to the survey limits set by the photometry.

The power of {\it spectral} classification in a multi-color survey depends both on the filter set used and the depth of the imaging, where the optimum choices are determined by the goal of the survey. If a survey aims at identifying only one type of object with characteristic colors, a tailored filter set can be designed. E. g., when looking exclusively for U-band dropouts \cite{Ste95}, the UGR filter set is certainly a very good choice. The performance of such a dropout survey depends mostly on the depth reached in the U-band, so the photon flux detection limit in U is the key figure. Also, number count studies are limited by the completeness limit in the filter of concern. Quasar search is very often done with color excess rules \cite{Haz90}, where the limit is given by the flux errors combined from two or three filters. E.g., the evolution of quasars between redshift 0 and 2.2 was established using the UV excess method \cite{SG83,BSP88}. At higher redshift quasars display rather star-like broad-band colors, motivating more advanced approaches like the selection of outliers in an n-dimensional color space \cite{WHO91}. 

If we now intend to focus different survey programs on a common patch of sky to maximise synergy effects from the various efforts, then we might as well combine the individual surveys into one that identifies every object, and avoid double work. Then we have to ask for a filter set which enables identifying virtually every object above some magnitude limit unambigously. In this case, the key number for the performance is the magnitude limit for a successful classification as needed for various science applications. If the classification takes all available color data into account, like template fitting procedures do, then the flux limit of a single filter is not the only relevant number, since the performance will depend to a large extent on the filter choice. This applies also for the estimation of multi-color redshifts, an idea dating back to Baum (1962), who used nine-band photoelectric data to estimate the redshifts of galaxy clusters.

Most multi-color surveys conducted to date obtained spectral information via broad-band photometry. They have been used e.g. to search for quasars or for high-redshift galaxies. However, they always needed follow-up spectroscopy to clarify the true nature of the candidates and to measure their redshift. The SLOAN Digital Sky Survey \cite{York00} is now the most ambitious project to provide a broad-band color database, on which the astronomical community might perform a large number of ``virtual surveys''.

So far, only very few survey projects make extensive use of medium-band and narrow-band photometry, e.g. the Calar Alto Deep Imaging Survey \cite{Mei98}. Surveys like CADIS with typically 10 to 20 filters are sampling the visual spectrum with a resolution comparable to that of low resolution imaging spectroscopy. CADIS fostered the development of a scheme for spectral classification, that distinguishes {\it stars}, {\it galaxies}, {\it quasars} and {\it strange objects}. Simultaneously, it assigns multi-color redshifts to extragalactic objects. 

Using 162 spectroscopic identifications Wolf et al. (2000a, henceforth paper II) have shown, that it is reliable for virtually all objects above the 10-$\sigma$ limits of the CADIS survey. Also, the photometric redshifts are accurate enough ($\sigma_z \approx 0.03$ for galaxies and $\sigma_z \approx 0.1$ for quasars around the 10-$\sigma$ limit), so that follow-up spectroscopy is not needed for a number of analyses, e.g. the derivation of galaxy luminosity functions \cite{Fri00}.

After this algorithm was developed for CADIS, it is now used for classification in additional projects. It provides multi-color redshifts in lensing studies of the cluster Abell 1689 \cite{Dye00}, aiming at determining the cluster mass after identifying cluster members and weakly lensed background objects. It is also employed for an ongoing widefield survey to search for high-redshift quasars, to provide multi-color redshifts for galaxy-galaxy lensing studies, to search for high-redshift galaxy clusters and to perform a census of L* galaxies at $z \approx 1$ \cite{Wolf00b}.

The purpose of this paper is to present our classification scheme and discuss the optimization of its use for optimum survey strategies. The statistical algorithm for the scheme is presented in Sect. 2 and our choice for the template libraries is detailed in Sect. 3. In Sect. 4 we report on simulations of a few competitive filter sets and their expected classification performance. We include an analytic discussion on the comparison of filter sets and conclude that medium-band surveys are altogether more powerful, even when being limited by available telescope time. Sect. 5 outlines a few real datasets using this classification and draws conclusions about the expected performance. Paper II demonstrates real CADIS data based on which we gained experience during the development of the scheme, and show, that the conclusions from the simulations compare well to the real dataset.

\section{The classification algorithm}

\subsection{General remarks on classification}\label{algor_general}

Generally speaking, classification is a process of pattern recognition which usually has to deal with noisy data. Mathematically, a classifier is a function, which is mapping a feature vector of a measured object characteristics onto a discriminant vector, that contains the object's likelihoods for belonging to the different available classes. Any classification relies on the feature space being chosen such that different classes cover different volumes and overlap as little as possible to avoid ambiguities.

If a survey is designed without class definitions in mind, it will be difficult to choose a set of measurable features for a tailored classification. Also, only {\it unsupervised} classifiers (= working without knowledge input) can be used to work on measured object lists. In this case, a classifier can find distinguishable classes, e.g. by cluster analysis. This process leads to a definition of new class terms which depends strongly on the visible features taken in account. 

For any classification problem, it is of great advantage, if class terms are defined {\it a priori} and encyclopedic knowledge is available about measurable features and their typical values. Then models of the classes representing this knowledge can be constructed to serve as an essential input to a {\it supervised} classifier (= using input knowledge as a guide). When selecting the features, two potential problems should be avoided: One is the use of well-known but hardly discriminating features, which will obviously not improve the classification but just increase the effort. The other is using features which are not well-known and therefore can easily cause mistakes in the classification. Especially, with high measurement accuracy this can lead to apparent unclassifiability when an object looks different than expected.

Two different types of class models can be distinguished depending on the uniqueness of the classification answer:
\begin{enumerate} 
\item In one type of models geometric rules are used to delimit sectors in feature space covered by the classes in competition. These models assign just one class to the measurement uniquely and definitely, which is the one containing the feature vector within its geometric limits. Effectively, the discriminant vector does not contain likelihood values in a statistical sense but instead a single entry '1' for the class decided on and zeros for the other classes (while nearest-neighbor classifications can define rather complicated boundary shapes in feature space, they also belong to this type). 

\item Another type are statistical class models rendered as likelihood functions which are defined across the entire feature space range. Only these provide discriminant vectors with relative likelihoods of class membership for an object, thereby following a ``fuzzy logic'' approach. 
\end{enumerate} 

While classes are discrete entities, a statistical classification can also work on continuous parameters. The discriminant vector then becomes a likelihood function of the parameter value. Based on this distinction {\it classification} problems can be considered as {\it decision} problems for discrete variables and {\it estimation} problems for continuous variables \cite{MC78a}. In either case, a definite statistical classification containes two consecutive steps: First, the discriminant vector is determined (see Sect.\,\ref{algor_step1}) and second, it is mapped either by decision to a final class or to a parameter estimate (see Sect.\,\ref{algor_step2}).

\subsection{Step 1: Determining discriminant vectors}\label{algor_step1}

We assume an object with $m$ features being measured by any device, thus displaying the feature vector $\vec q = (q_1, \ldots q_m)$. We consider $n$ classes $c_1, \ldots c_n$ as a possible nominal interpretation and denote the likelihood of this object to belong to the class $c_i$ as $p(c_i|\vec q)$. A true member of class $c_i$ has an {\it a-priori} probability of displaying the features $\vec q$ given by $p(\vec q|c_i)$. 

Initially, we assume a simple case of uniquely defined class models, where all members of a single class $c_i$ have the same intrinsic features $\vec q_{c_i}$, so that any spread in measured $\vec q$ values arises solely from measurement errors. Assuming a Gaussian error distribution for every single feature, it follows \cite{MC78b}, that
\begin{eqnarray}
	p(\vec q | c_i) = C \exp \left(-\frac{1}{2} (\vec q-\vec q_{c_i}) 
				V^{-1}_k (\vec q-\vec q_{c_i})^t \right) ~,
\end{eqnarray}

where $(\vec q-\vec q_{c_i})$ is the measurement error in case the object does belong to $c_i$ and $(\vec q-\vec q_{c_i})^t$ is its transposed version. Each feature $q_k$ is measured with its own error variance $\sigma^2_k$, which are the diagonal elements in the variance-covariance matrix $V$. If all the features are statistically independent, the off-diagonal elements vanish. The normalisation factor $C$ is
\begin{eqnarray}
	C = \frac{1}{\sqrt{(2 \pi)^m |V_k|}} ~.
\end{eqnarray}

As contained in the discriminant vector, the likelihood for an object observed with $\vec q$ to belong to class $c_i$ is then
\begin{eqnarray}\label{prelativ}
	p(c_i | \vec q) = p(\vec q | c_i) / \sum^n_{l=1} p(\vec q | c_l) ~.
\end{eqnarray}

However, in realistic cases the classes themselves are extended in feature space and their volume might have rather complicated shapes. In the spirit of Parzen's kernel estimator \cite{Par63} the extended class $c_i$ can be represented by a dense cloud of individual uniquely defined (point shape) members $c_{ij}$. Every member accounts for some a-priori probability to display $\vec q$, given as $p(\vec q|c_{ij})$, just as if it were a ``class'' on its own. The complete class $c_i$ is now rendered as a superposition of its $N_i$ members and adds up to a total probability of
\begin{eqnarray}\label{pqci}
	p(\vec q | c_i) = \frac{1}{N_i} \sum_j p(\vec q | c_{ij}) ~.
\end{eqnarray}

In an {\it estimation} problem the probability functions have the same form, except for changes in the notion: $\theta$ denotes the parameter to be estimated, and ideally the class model $c_i$ had a continuous shape covering the range of expected values. The discriminant vector would then be a function $p(\theta|\vec q)$. Again, the class model can be approximated by a discrete set of members sampling the $\theta$ range of interest at sufficient density.

The astronomical application discussed in this paper poses a mixture of decision and estimation problems which can be realized simultaneously with a unified approach: The decision may choose from the three classes $c_1 =$ stars, $c_2 =$ galaxies and $c_3 =$ quasars, and an estimation process takes care of the parameters redshift and different spectral energy distributions (SED). The internal structure of every class $c_i$ is then spanned by its individual parameter set $\vec \theta_i = \{\theta_{i\vec j} \}$, either following a grid design or being unsorted if no parameter structure is needed. 

If one chooses to approximate the spatial extension of a class by a dense grid sampling discrete parameter values, two problems are solved at once: on the one hand, an internal structure is present for estimating parameters, and on the other hand, the class is well represented for calculating its total probability $p(\vec q|c_i)$. Altogether, the probability function with internal parameters $\theta_{i\vec j}$ being resembled by class members $c_{i \vec j}$ is then
\begin{eqnarray}
	p(\vec q | \theta_{i \vec j}) = 
			C \exp \left(-\frac{1}{2} (\vec q-\vec q(\theta_{i \vec j})) 
				V^{-1}_k (\vec q-\vec q(\theta_{i \vec j}))^t \right)
\end{eqnarray}

with the total probability for class $c_i$ being
\begin{eqnarray}\label{pqci2}
	p(\vec q | c_i) = \frac{1}{N_i} \sum_{\vec j} p(\vec q | \theta_{i\vec j}) ~,
\end{eqnarray}

and the equation for the class likelihood function still being
\begin{eqnarray}
	p(c_i | \vec q) = p(\vec q | c_i) / \sum^n_{l=1} p(\vec q | c_l) ~.
\end{eqnarray}

Based on these probability functions the classification can perform a decision between object classes and estimations of redshift and other object parameters at once. Two different analyses are integrated into one paradigm and calculated efficiently by evaluating the same probability density function.

\subsection{Step 2: Decision and estimation}\label{algor_step2}

Decision rules are functions mapping a discriminant vector $\vec p(c_i|\vec q)$ to a decision value $d$. The value $d_i$ denotes a decision in favor of class $c_i$, i.e. the object displaying features $\vec q$ is then assumed to belong to this class. The most simple decision rule is the maximum likelihood (ML) scheme, which decides for the one class with the highest likelihood $p$. In case of two classes existing this means
\begin{eqnarray}
	{\mbox{if \quad $p(c_1 | \vec q) > p(c_2 | \vec q)$ \quad, then}\quad d_1 \atop
	 \mbox{if \quad $p(c_1 | \vec q) < p(c_2 | \vec q)$ \quad, then}\quad d_2} ~.
\end{eqnarray}

A more compact notion for the same rule is
\begin{eqnarray}
	p(c_1 | \vec q) \begin{array}{c} d_1 \\ ^>_< \\ d_2 \end{array} p(c_2 | \vec q) ~.
\end{eqnarray}

Depending on the purpose of the classification tailored improvements can be made to this rule. The probability of error (PoE) method, e.g., attempts to minimize the rate of misclassifications by including the a-priori-probability for observing a member of a given class. Following Bayes theorem these ``priors'', denoted $P(c_1)$ and $P(c_2)$, are just the relative abundance of the class in the whole sample. The PoE decision rule is then 
\begin{eqnarray}
	p(c_1 | \vec q) P(c_1) \begin{array}{c} d_1 \\ ^>_< \\ d_2 \end{array} 
						p(c_2 | \vec q) P(c_2) ~,
\end{eqnarray}

which causes somewhat ambiguous objects to be preferentially classified as belonging to the more common class. Rare objects are then less likely to be found at all, but the overall performance of the classifier improves. A general approach uses any type of priors for trimming the classification towards specific goals, so every decision rule compares the likelihood ratio $\Lambda$ with a threshold $T$ and follows the form (with $T=1$ for ML decision)
\begin{eqnarray}
	\Lambda (\vec q) = \frac{p(c_1 | \vec q)}{p(c_2 | \vec q)}  
			\begin{array}{c} d_1 \\ ^>_< \\ d_2 \end{array} T
\end{eqnarray}

Estimation rules are functions mapping a discriminant vector $\vec p(\theta|\vec q)$ to an estimated value $\tilde\theta$. The most simple estimation rule is again the maximum likelihood (ML) rule, which chooses the one paramter value with the highest likelihood $p$, i.e., the ML estimator is given by
\begin{eqnarray}
	p(\tilde\theta_{ML} | \vec q) \geq p(\theta | \vec q) \qquad \forall \theta ~.
\end{eqnarray}

The Bayesian approach can also be applied to continuous variables, whereas one special case is of particular interest: if the error distribution of the feature measurement is Gaussian, and if the goal is to minimize the variance of the true estimation error, then the optimum estimation rule can be derived analytically \cite{MC78b}. This minimum error variance (MEV) estimator is given by
\begin{eqnarray}
	\tilde\theta_{MEV} = \frac{\int \theta p(\theta | \vec q) P(\theta) \,d\theta}
			{\int p(\theta | \vec q) P(\theta) \,d\theta} ~,
\end{eqnarray}

and it is equivalent to interpreting the discriminant vector as a statistical ensemble and determining the mean of the distribution. It is also dubbed mean square estimator or conditional mean estimator. Note that, if $p(\theta|\vec q)$ is symmetric in $\theta$ and unimodal, the MEV estimator is {\it identical} to the ML estimator.

\subsection{Application to astronomical multi-color surveys}\label{algor_applic}

Deep extragalactic surveys usually contain mostly galaxies, fewer stars and a tiny fraction of quasars, with relative numbers on the order of 100:10:1. A survey at galactic latitudes above {\bf $|b| \ga 50\degr$} with a limiting magnitude of $R=23$ and an area of 1\,$\sq\degr$, e.g., should contain roughly 30000 galaxies \cite{Met95}, some 3000 to 6000 stars \cite{BS81,Phl00}, and about 400 quasars including Seyfert-1 galaxies \cite{HS90}. Any classification would ideally be capable of distinguishing all three classes of objects. Only in surveys, which do not care about the rare quasars, their class could be dropped and the classification needed to separate only stars from galaxies.

In addition to the class itself, plenty of physical parameters could potentially be recovered from an object's photometric spectrum. Most importantly, we would like to determine redshift estimates for galaxies and quasars. In addition, the spectral energy distribution of galaxies contains information about their star formation rate and the age of their stellar populations. A photometric spectrum of sufficiently high spectral resolution can even allow to estimate the intensity of emission-lines. Finally, the spectra of stars tell mostly their effective temperature, but also their metallicity and their surface gravity.  

The literature provides abundant knowledge of spectral properties for all three object classes. Synthetic photometry can use published spectra together with efficiency curves from the survey filter set in order to obtain predicted colors of objects. Sometimes, model assumptions are needed to fill in data gaps present in the literature, which could either be gaps on the spectral wavelength axis or gaps on physical parameter ranges, e.g. star-formation rate. Eventually, systematic multi-color class models can be calculated from published libraries covering various physical parameters. These can serve for later comparison with observed data. Therefore, we decided to build a statistical classification based on published spectral libraries and a limited number of model assumptions (see Sect.\,3).

In a multi-color survey the dominant information gathered are the object fluxes in the different filters. We decided to use the color indices as an input to the classification rather than the fluxes themselves, which eliminates one dimension from the problem by omitting the need for any flux normalisation, that remains as an additional fit paramter in template fitting procedures. It will be shown in Sect.\,\ref{algor_equivalent}, that the color-based approach is equivalent to the flux-based one under certain constraints. 

Morphological information is typically also available to some extent and can be included in the classification based on the assumption that only galaxies are capable of showing spatial extent. But this should be done carefully, since luminous host galaxies can render quasars as extended. Also, if the image quality varies across the observed field, the morphological analysis is of limited use for not clearly extended sources.

We define the color $q_{g-h}$ as a magnitude difference between the flux measurements in two filters $F_g$ and $F_h$:
\begin{eqnarray}
	q_{g-h} = m_g-m_h = -2.5\log\frac{F_g}{F_h} ~.
\end{eqnarray}

Obviously, the color system depends on the filter set chosen and also on the flux normalisation used. As long as the flux errors are relatively small, the linear approximation of the logarithm can be used to express magnitude errors as $\sigma_{m_i} \approx \sigma_{F_i} / F_i$, so that the error of the color is
\begin{eqnarray}
	\sigma_{q_{g-h}} = \sqrt{\sigma^2_{m_g} + \sigma^2_{m_h}} \approx \sqrt{
		 (\sigma_{F_g} / F_g)^2 + (\sigma_{F_h} / F_h)^2} ~.
\end{eqnarray}

Since the likelihoods determined for the classification depend sensitively on the colors $\vec q$ {\it and} their errors $\vec \sigma_q$, both values must be carefully calibrated. If any color offset is present between measurement and model, the classification will go wrong systematically. If errors are underestimated, the likelihood function could focus on a wrong interpretation, rather than including the full range of likely ones. Overestimated errors will obviously diffuse the likelihoods and give away focus which is originally present in the data. The approximation of errors as presented will only work well with flux detections of at least 5$\sigma$ to 10$\sigma$, but at lower levels the classification is likely to fail anyway, so we ignore this concern.

Given $\vec q$ and $\vec \sigma_q$ a measured object is represented by a Gaussian error distribution rather than a single color vector. If colors are measured very accurately and the object is rendered as a narrow distribution, it could possibly fall between two grid steps of a discrete class model and ``get lost'' for the classification. In this case low likelihoods would be derived despite the spatial proximity of object and model in terms of metric distance. The likelihood function would appear not much different from that of a truely strange object residing off the class in an otherwise empty region of color space. In technical terms, the classification would violate the {\it sampling theorem} \cite{Jae91}, and the probability functions would not be invertible any more.

For discrete class models the sampling theorem requires that every measurement falling inside the volume of a model should ``see'' at least two model members inside of its Gaussian core. Due to practical limitations of computing time and storage space, it does not make sense to develop discrete models with virtually infinite density accounting for arbitrarily sharp measurements. Also, for measurements with low photon noise the dominant source of error will be the limited accuracy of the color calibration. 

The solution to the problem is then to design the discrete model with the achievable measurement accuracy in mind, and to smooth the discrete model into a continuous entity by convolving its grid with a continuous function that is wide enough to prevent residual low-density holes between the grid points. A sensible smoothing width would just fulfill the sampling theorem, i.e. the smoothing function should roughly stretch over a couple of discrete points. As a result, even an extremely sharp measurement will be covered by the model and classified correctly. 

Higher resolution would only increase the computational efforts while lower resolution would ignore information which is present in the data and therefore potentially worsen the classification. From a different point of view, one could leave the discrete model unchanged and claim the data to have larger effective errors by including the calibration errors thereby limiting the width of the Gaussian data representation to a lower threshold, which will always ensure the sampling theorem on the discrete grid anyway.

Both approaches are mathematically identical, if one chooses to represent the calibration errors as well as the smoothing function by a Gaussian. Due to the symmetry of the Gaussian function, convolving the discrete grid or convolving the error distribution of the data yields the same result. The choice of the Gaussian is computationally very efficient, because the convolution  of the Gaussian measurement with the Gaussian calibration error results in another Gaussian of enlarged width. As mentioned in Sect.\,\ref{calib_data} and discussed in paper II, a survey in the visual bands can be calibrated with a relative accuracy on the order of 3\% between the different filters. Therefore, we decide to apply a $0\fm03$-Gaussian as a smoothing function.

In summary, we apply the formalism presented in Sect.\ref{algor_step1} in the following way: the errors $\sigma_{q_i}$ of the colors $q_i$ are convolved with the smoothing $0\fm03$-Gaussian and as a result the effective errors are
\begin{eqnarray}
	\sigma^2_i = \sigma^2_{q_i} + (0\fm03)^2 ~.
\end{eqnarray}

For simplicity, we assume the individual colors to be uncorrelated, which is actually not true for filters sharing spectral regions in their transmission. The variance-covariance matrix then becomes diagonal
\begin{eqnarray}
	V_k = \left( \begin{array}{cccc} \sigma^2_1 & & & \\ & \sigma^2_2 & & \\ 
		& & \ddots & \\ & & & \sigma^2_m \end{array} \right) ~,
\end{eqnarray}

and the probability function turns into
\begin{eqnarray}
	p(\vec q | c_i) = \frac{C}{N_i} \sum_{j=1}^{N_i} \exp \left(-\frac{1}{2} 
		\sum_{k=1}^m \left( \frac{q_l - q_{c_{ij},k}}{\sigma_k} \right)^2 \right) ~.
\end{eqnarray}

Based on the three object classes discussed the likelihood function is
\begin{eqnarray}
	p(c_i | \vec q) = \frac{p(\vec q | c_i)}{p(\vec q | c_{stars}) + 
			p(\vec q | c_{galaxies}) + p(\vec q | c_{quasars})} ~.
\end{eqnarray}

Considering three classes implies that extremely faint objects with large errors get average probabilities of 33\% assigned for all classes. In general applications, we use a decision rule for an object seen as $\vec q$, which requires that one class is at least three times more probable than the other two classes put together, i.e.:
\begin{quote}
If there is one class with $p(c_i|\vec q) > 0.75$, then we assign this class to the object, but if all classes have likelihoods below 0.75, we call it {\it unclassifiable}.
\end{quote}

For the detection of unusual objects, we look at the color distance of an object to the nearest member of any class model to derive a statistical consistency with the class. The value of this consistency depends on the different color variances and can be calculated from $\chi^2$-statistics. Lacking an analytic expression we use $\chi^2$-tables \cite{AS72} to evaluate the statistical consistency between class and object. In practice, the resulting $\chi^2$-values need to be normalised to a plausible scale, since the raw values obtained are enlarged artificially due to the discrete sampling of the library and cosmic variance. We use the following operative criterion for the selection of unusual objects:
\begin{quote}
If an object is inconsistent at least at a confidence level of 99.73\% (i.e. 3$\sigma$ in case of a normal distribution) with all members of all classes, then we call it {\it strange}.
\end{quote}

Strange objects can formally be classifiable, if the likelihoods still prefer a certain class membership. They have either intrinsically different spectra without counterparts in the class models, or they are reduction artifacts, e.g. when neighboring objects affect their color determination, and this is not taken into proper account for the error calculation.

Apart from the rather trivial ML estimator, we use the MEV estimator to obtain redshifts and SED parameters of galaxies and quasars. Their class models are designed as regular grids (see Sect.\,3) with members $c_{ij}$ residing at redshift $z_{ij}$. The MEV estimator for the redshift is then
\begin{eqnarray}
	\langle z \rangle _{MEV} = \frac{\sum_j z_{ij} p(\vec q | c_{ij})}
				{\sum_j p(\vec q | c_{ij})} ~.
\end{eqnarray}

It is applied to the class models for galaxies and quasars independently and for each class interpretation an independent redshift estimate is obtained. There is also an assessment for the likely error of the $z$ estimate given by the variance of the distribution $p(\vec q|z)$:
\begin{eqnarray}
	\sigma^2_z = \frac{\sum_j (z_{ij}-\langle z \rangle _{MEV})^2 p(\vec q | c_{ij})}
				{\sum_j p(\vec q | c_{ij})} ~.
\end{eqnarray}

This estimation scheme would be sufficient, if models had a rather simple shape in color space, i.e. if color space and model parameter space could easily be mapped onto each other. In fact, the class model for galaxies and particularly the one for quasars can have very complicated folded shapes in color space, so that the distribution $p(\vec q|z)$ can have a correspondingly complicated structure that is not at all well described by mean and variance. 

Therefore, we distinguish three cases: unimodal (single peaked), bimodal (double peaked) and broad distributions. In unimodal cases $\langle z \rangle _{MEV}$ and $\sigma_z$ are appropriate reductions of $p(\vec q|z)$. In bimodal cases we split the redshift axis in two intervals delimited at $\langle z \rangle _{MEV}$ and obtain two alternative unimodal solutions with relative probabilities given by the $p$ sums in the two intervals. If the distribution is so broad, that it starts to resemble a uniform distribution, $\langle z \rangle _{MEV}$ approaches the mean $z$ value of the model and $\sigma_z$ approaches $\sqrt{1/12} (z_{max}-z_{min})$. In order to keep our statistics clean from such mean redshift contaminants, we cut off the estimator at some uncertainty:
\begin{quote}
If an object has $\sigma_z > 1/8 (z_{max}-z_{min})$, then we ignore the MEV estimate and call its redshift {\it uncertain}.
\end{quote}

In particular, it is possible, that an object has a bimodal distribution with one peak (result) and one broad ({\it uncertain}) component. In the following, we denote this extended scheme of MEV estimates accounting for possible bimodalities as our MEV+ estimate. In Sect.\,\ref{mlvsmev} we will compare the performance of all three estimators, ML versus MEV and MEV+.

An effort was made to implement a classification code optimized for short computing time. The use of precalculated class models eliminates any synthetic photometry from a typical fitting procedure. Furthermore, the use of colors instead of fluxes eliminates the need for finding a flux normalisation. In terms of CPU time, the classification of one object contains mainly the calculation of the probability $p(\vec q|c_{ij})$ for every class member, which involves first adding up all $\sigma^2_i$-scaled squared color differences and second evaluating an exponential function of the resulting sum that is already a measure of strangeness. Summing up the $p(\vec q|c_{ij})$ to obtain class likelihoods and deriving mean and variance of the internal class parameters should take less time than calculating the probability density function, if more then ten color axes are taken into account. With class models containing about 50000 members and 13 colors, the full classification of one object takes about 0.3\,sec when running on a 200\,MHz Ultra Sparc CPU inside a SUN workstation. Since different survey applications might require different sample selection schemes, we decided to calculate and store discriminant vectors for all objects and select subcatalogs for further analysis later.

\subsection{Equivalence of flux-based and color-based classification}\label{algor_equivalent}

We now show, that the color-based classification yields the same best fit as a flux based template fitting algorithm. Lanzetta et al. (1996), e.g., calculate a likelihood function depending on redshift $z$, a spectral energy distribution and a flux normalisation parameter $A$, following the form:
\begin{eqnarray}
	L_{model} = \exp \left(-\frac{1}{2} 
		\sum_k \left( \frac{F_{k,obs} 
		- A \tilde F_{k,model}}{\sigma_{F_k}}\right)^2 \right) ~.
\end{eqnarray}

Basically, the likelihood determination relies on the squared photometric distance $d$ between observation and model, resulting from the flux differences $\Delta F_k$ in each filter:
\begin{eqnarray}\label{fluxTF}
	d^2 = \sum_k^n \chi_k^2   \quad \mbox{with} \quad
		\chi_k = \frac{F_{k,obs} - F_{k,model}}{\sigma_{F_k}} 
			= \frac{\Delta F_k}{\sigma_{F_k}} ~.
\end{eqnarray}

In the color based approach there are $n-1$ color indices contributing distance components and we assume the single constraint, that there is one particular {\it base filter} approximately free of flux errors, e.g. a deeply exposed broad-band filter. The color indices are made by comparing any filter to this {\it base filter} ensuring optimum errors for the colors. In this scheme, any errors in the relative calibration are absorbed into the color indices. Therefore, it is very important, that the {\it base filter} is not wrongly calibrated with respect to the other wavebands, since the error would spread into the entire vector of color indices.

We then look only at a range of good fits, and do not mind rather crude $\chi$-approximations for relatively bad fits which are anyway ruled out as solutions. Also, we consider only measurements with $\sigma_{F_k} / F_k \la 0.2$, which allows the assumption of Gaussian color errors and a linear approximation of the logarithm. The distance components are:
\begin{eqnarray}
	\chi_k & = & \frac{(m_k-m_{base})_{obs}-(m_k-m_{base})_{model}}{\sigma_{m_k-m_{base}}}
								\nonumber \\
		& = & 2.5 \frac{log (F_k/F_{base})_{obs} - log (F_k/F_{base})_{model}}
			{\sqrt{(\sigma_{F_k}/F_k)^2+(\sigma_{F_{base}}/F_{base})^2}} ~.
\end{eqnarray}

Using the terms $\Delta F_k$ and $\sigma_{F_{base}}/F_{base} \approx 0$, we obtain
\begin{eqnarray}
	\lefteqn{\chi_k \approx 2.5 \frac{F_k}{\sigma_{F_k}} \cdot} \nonumber \\
		& & \left\{ log \left(1+\frac{\Delta F_k}{F_{k,model}} \right) 
	 	- log \left(1+\frac{\Delta F_{base}}{F_{base,model}} \right) \right\} ~.
\end{eqnarray}

Expanding the logarithm for $\Delta F_k/F_k \ll 1$, we get
\begin{eqnarray}
	\chi_k & \approx & \frac{F_k}{\sigma_{F_k}} \left\{ \frac{\Delta F_k}{F_{k,model}} -  
			\frac{\Delta F_{base}}{F_{base,model}} \right\} \nonumber \\
	       & \approx & \frac{\Delta F_k}{\sigma_{F_k}} 
		+ \frac{\Delta F_k}{\sigma_{F_k}} \frac{\Delta F_k}{F_{k,model}}
		- \frac{\Delta F_{base}}{\sigma_{F_k}} \frac{F_k}{F_{base,model}} ~.
\end{eqnarray}

The first term is typically on the order of 1, while the second term is on the order of $\sigma_{F_k}/F_k \ll 1$ and the third one of $\sigma_{F_{Base}}/\sigma_{F_k} \ll 1$. Therefore, the last two terms can be dropped and the expression for $\chi_k$ reduces to
\begin{eqnarray}
	\chi_k \approx \frac{\Delta F_k}{\sigma_{F_k}} ~,
\end{eqnarray}

which is identical to the expression used in the flux template fitting method shown in Eq. \ref{fluxTF}.

\subsection{System of color indices}

In the previous section, we had discussed the relevance of a common {\it base filter} for the various color indices, which is supposed to have relatively small flux errors in order to keep the color errors as low as possible. Our multiband survey applications usually involve a smaller number of broad bands as well as a larger number of medium-band observations. For these, we decided to form color indices from broad bands neigboring on the wavelength axis, i.e. U--B, B--V, V--R and R--I, which we assume to be the optimum solution for comparably deep bands. Each of the shallower medium bands we combine with the most nearby broad-band in terms of wavelength, which then serves as a {\it base filter} for the medium-band color indices, e.g. B--486 or 605--R, where letters denote broad bands and numbers represent the central wavelength of medium-band filters measured in nanometers. 

In terms of flux template fitting, this scheme of color indices means, that we use a few deep broad bands to fit the global shape of the SED, and then use a few groups of medium bands around each deep broad-band to fit the smaller-scale shape locally. The $\chi^2$-values of the global fit and the several local fits are then just added up to the total $\chi^2$. This scheme has a particular advantage over a solely global flux fitting: the local fits can well trace spectral structures, even if the global distribution of the object differs from the template (e.g. as it could be caused by extinction). Therefore, it is not too dependent on accurate global template shapes and it can use the ability of the medium bands to discriminate narrow spectral features for a more accurate classification. Of course, this advantage vanishes immediately for a pure broad-band survey, where local structures in the spectrum are not traced, and therefore no local fits are available for the $\chi^2$-sum.

\section{The classification libraries}

\begin{figure*}
\centerline{\hbox{
\psfig{figure=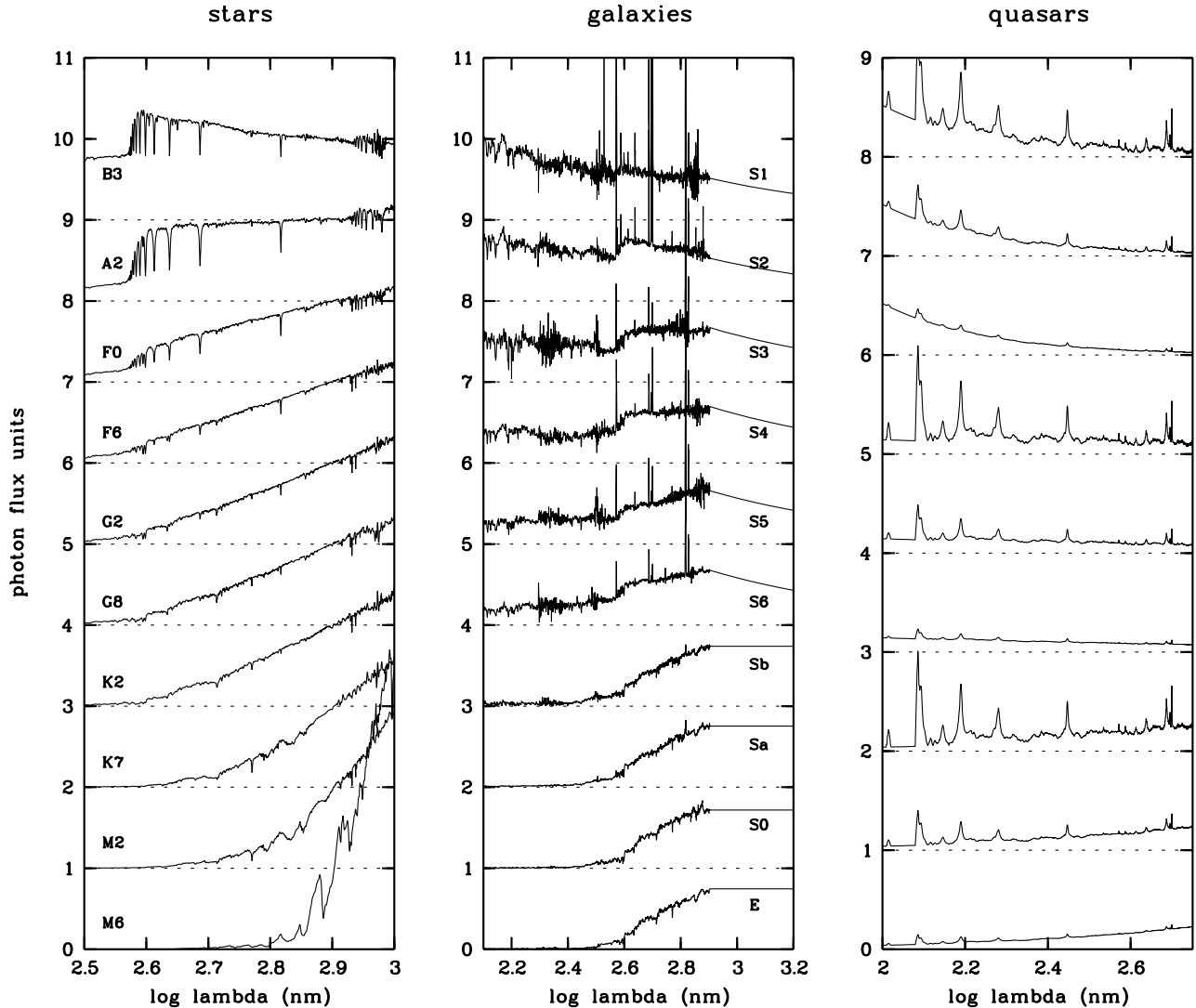,angle=270,clip=t,width=17cm}}}
\caption[ ]{This diagram shows a few selected spectra from our template libraries. The shown wavelength scale runs from 315\,nm to 1000\,nm for stars (left), from 125\,nm to 1600\,nm for galaxies (center) and from 100\,nm to 550\,nm for quasars (right). The flux is $\lambda f_\lambda$ in units of photons per nm, time intervall and sensitive area and offset by one unit per step within a class. The flux scale is normalised to unity at 800\,nm for stars, arbitrary for galaxies, and normalised to 0.2 at 250\,nm for quasars. The stellar templates are taken from Pickles (1998), the galaxy templates from Kinney et al. (1996) and quasar templates are modelled after Francis et al. (1991). The quasar diagram shows nine spectra with three different spectral indices (-2.0, -0.6, +0.8) and three different relative emission-line intensities (0.6, 2.1, 5.7). \label{libs}}
\end{figure*}

We assembled the color libraries from intrinsic object spectra assuming no galactic reddening. Clearly these libraries can only be sufficient when observing fields with low extinction and little reddening. Usually, such fields are chosen for deep extragalactic surveys and the CADIS fields in paticular were carefully selected to show virtually no IRAS 100\,$\mu$ flux (below 2\,MJy/sterad), so we expect ``zero'' extinction and reddening there. When applying this color classification to fields with reddening, the libraries would have to be changed accordingly. 

Obviously, the libraries should contain a representative variety of objects, but still they can never be assumed to cover a complete class including all imaginable oddities. When classes are enlarged to cover as many odd members as possible, there is a trade-off to be expected between classifying the odd ones right, and introducing more spatial overlap between the classes in general, i.e. introducing more confusion among normal objects. The spectral libraries we employ are partly based on observations only and partly mixed with model assumptions. Our particular choice of libraries is founded on experience we gained within the CADIS survey, where we found several other published templates to be less useful.

\subsection{The star library}

For the stars, we picked the spectral atlas of Pickles (1998), that contains 131 stars with spectral types ranging from O5 to M8. It covers different luminosity classes but concentrates on main sequence stars, and it also contains some spectra for particularly rich metallicities. For the surveys in consideration, very young and very luminous stars should not be expected, but we include the entire library nevertheless (see Fig.\,\ref{libs}). Stars later than M8 are missing in the library, but they do show up in deep surveys like CADIS \cite{Wolf98}. These objects are interesting on their own, of course, but they are so rare, that a couple of misclassifications do not hurt the statistics on other objects.

In earlier stages of the CADIS survey, we reported using the Gunn \& Stryker (1983) atlas of stellar spectra (see e.g. Wolf et al. 1999), which has a number of disadvantages compared to the new work by Pickles. Our impression is that the Pickles spectra have a better calibration in the far-red wavelength range and are less affected by noise there. Especially, broad absorption troughs in M stars are rendered more accurately in the Pickles templates, which can be quite relevant for medium-band surveys. Also, they cover the NIR region and, e.g., the entire CADIS filter set all the way out to the K$^\prime$ band, thereby omitting the need for homemade extrapolations. Since it contains two different metallicity regimes, it covers the range of possible stellar medium-band colors better than the Gunn \& Stryker atlas, most notably among M stars for colors sensitive to their deep absorption features and, e.g., among K stars for colors probing the Mg I absorption.

The atlas is not structured as a regular grid in the stellar parameters and we consider the resulting color library an unsorted set without internal structure. If variations in dust reddening are to be expected within the field as in the case of Galactic stellar observations, this effect should be treated as an additional parameter in the library.

For multi-color surveys aiming specifically at Galactic stars, one would ideally like to have a library organized as a regular grid in effective temperature, surface gravity and metallicity, which could, e.g., be derived from model atmospheres. Such a fine classification is not needed for extragalactic surveys, where the focus is on galaxies and quasars. We gained some experience with the stellar spectra from the model grid by Allard (1996), but we decided not to use it, since the overall colors seemed to be better matched by the Pickles library.

\subsection{The galaxy library}

The galaxy library is based on the template spectra by Kinney et al. (1996). These are ten SEDs averaged from integrated spectra of local galaxies ranging in wavelength from 125 nm to 1000 nm. The input spectra of quiescent galaxies were sorted by morphology beforehand to result in four templates called E, S0, Sa and Sb. The starburst galaxies were sorted by color into six groups yielding six more templates called SB6 to SB1. Based on the observation, that color and morphology of galaxies correlate, this template design seems reasonable. This way the classification can indirectly measure morphology of galaxies via their SED, at least as far as the locally determined color-morphology relation holds at higher redshift.

The templates contain a very deep unidentified absorption feature around 540 nm, which we supposed to be an artifact of the data reduction and eliminated. We left the abundant structures in the UV unchanged, although some of them might be noise and we do not know how to interprete them. We modelled a near-infrared addition heuristically by a simple law consistent with the $I-K^\prime$-colors of a sample of galaxies with known spectroscopic redshifts (see paper II). Using this addition, we extended the spectra out to 2500 nm, and actually replaced the spectrum starting from 800 nm to eliminate the noise in the templates redwards of 800 nm (see Fig.\,\ref{libs}). Quiescent galaxies were extended according to $f_\nu \sim \nu^{-1}$, while starburst galaxies seemed most consistent with an extension of $f_\nu \sim \nu^{-1/3}$.

We consider the templates to form a one-dimensional SED axis of increasingly blue galaxies and fill in more templates to obtain a dense grid of 100 SEDs. Our interpolation is done linearly in color space, and the number of filled-in SEDs is chosen such, that the color space is filled rather uniformly. The new SEDs are denominated as numbers from 0 to 99, where the ten original SEDs used for the interpolation reside at the following numbers:

\begin{center}E - S0 - Sa - Sb - S6 - S5 - S4 - S3 - S2 - S1\end{center}
\begin{center}0 - 15 - 30 - 45 - 75 - 80 - 85 - 90 - 95 - 99\end{center}

Internal reddening is considered an important effect for the colors of galaxies and especially common among later types. While trying to account for it, we realized that its effect is merely one of shifting the zeropoint in the SED and hardly one of changing the redshift estimates. If we did introduce an independent reddening parameter, it would be almost colinear with the SED axis itself. Therefore, we opted for using the templates as determined from real galaxies and provided by Kinney et al. (1996), since they probably contain already a typical distribution of reddened objects. Due to our scheme of SED interpolation, we can still classify galaxies, which are reddened more or less than usual.

We also tried to change the SED interpolation scheme by relocating the templates to different SED numbers, which did not seem to improve the results. The color library was calculated for 201 redshifts ranging in steps of $\Delta z = 0.01$ from $z=0$ to $z=2$, finally containing $201 \times 100$ members. We did not intend to go beyond a redshift of 2, since our survey applications have typically not become deep enough, yet, to see such objects in useful numbers. 

The main shortcoming of this library is that the 1-dimensional SED allows no variation in emission-line ratios independent of the global galaxy color. Since medium-band filters can contain strong emission-line signals from faint galaxies, an observed emission-line ratio detected by two suitably located filters can be in disagreement with the global SED traced by all other filters. Since especially the CADIS filters are placed to deliver multiple detections of emission lines at several selected redshifts, some degradation in real performance could be expected with respect to the simulation (see paper II).

\subsection{The quasar library}

\begin{figure}
\centerline{\hbox{
\psfig{figure=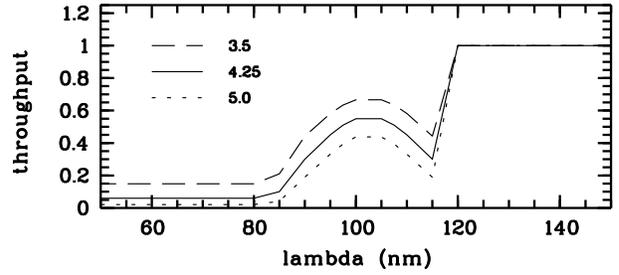,angle=270,clip=t,width=8cm}}}
\caption[ ]{For the quasars we assumed a throughput function for the Lyman-$\alpha$ forest which we derived from a visual inspection of quasar spectra published by Storrie-Lombardi et al. (1996). The scale of this function depends on redshift and is shown for $z=3.5$, $4.25$ and $5.0$. \label{qput}}
\end{figure}

The quasar library is designed as a three-component model: We add a power-law continuum with an emission-line contour based on the template spectrum by Francis et al. (1991), and then apply a throughput function accounting for absorption bluewards of the Lyman-$\alpha$ line. We modeled a throughput function $T_0$ after visually inspecting spectra of $z\approx 4$-quasars published by Storrie-Lombardi et al. (1996), and keep its shape constant (see Fig.\,\ref{qput}) while varying its scale to follow the increasing continuum depression $D_A$ towards high redshift. Using data from Kennefick (1996) and Storrie-Lombardi et al. (1996) as a guideline, we arrived at
\begin{eqnarray}
	T (z) = T_0^{(z/4.25)^2} ~.
\end{eqnarray}

\begin{figure}
\centerline{\hbox{
\psfig{figure=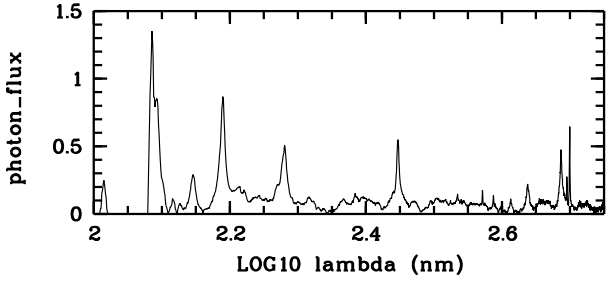,angle=270,clip=t,width=8cm}}}
\caption[ ]{The quasar library is based on an emission line contour taken from the quasar template spectrum by Francis et al. (1991). The wavelength scale runs from 100\,nm to 550\,nm and the flux is $\lambda f_\lambda$ in units of photons per nm, time intervall and sensitive area (arbitrary units). \label{qemi}}
\end{figure}

The intensity of the emission-line contour was varied only globally, i.e. with no intensity dispersion among the lines. As long as typically only one medium-band filter is brightened by a prominent emission line, the missing dispersion should not affect the classification (see Fig.\,\ref{qemi}). For the intensity factor relative to the template, $e^\epsilon$, ten values were adopted ranging in steps of $\Delta \epsilon = 0.25$ from $\epsilon = -0.5$ to $\epsilon = 1.75$ on a logarithmic scale, which is roughly 0.6 times to 5.7 times the template intensity. Originally, we tried a range from 0.3 times to 2.7 times, but the first twenty quasars found in CADIS contained mostly strong lines, which are better represented by the current limits. 

The slope of the power-law continuum $f_\nu \sim \nu^\alpha$ was varied in 15 steps of $\Delta \alpha = 0.2$ ranging from $\alpha = -2.0$ to $\alpha = 0.8$. The library was calculated for 301 redshifts ranging in steps of $\Delta z = 0.02$ from $z=0$ to $z=6$, finally containing $301 \times 15 \times 10 = 45150$ members. As a future improvement one could imagine the inclusion of Seyfert I galaxies with nuclei of rather low luminosity, i.e. spectra coadded as a superposition of a host galaxy spectrum with a broad-line spectrum for the nucleus.

\subsection{Calculation of color libraries}

As a first step, the spectral libraries were transformed into color index libraries representing precisely the set of filters and instruments in use. The use of precalculated filter measurements rather than fully resolved flux spectra removes any computationally expensive calculations for synthetic photometry from the process of classifying the object list. The use of color indices omits the needs for any flux normalisation, further speeding up the classification. A list of $\sim 10^4$ objects and $\sim 10$ colors can be classified within a couple of hours on a SUN Enterprise II workstation even when using $\sim 10^5$ templates. 

For best results it is required that the color libraries are calculated for an instrumental setup resembling precisely the observed one, i. e. the synthetic photometry calculation has to take every dispersive effect into account. We decided to use photon flux colors derived from the observable object fluxes, averaged over the total system efficiency of each filter and assuming an average atmospheric extinction.

\begin{figure*}
\centerline{\hbox{
\psfig{figure=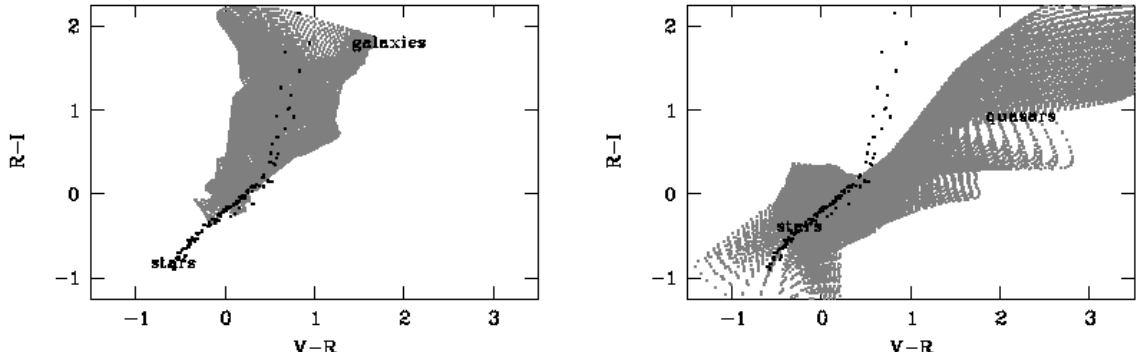,angle=270,clip=t,width=15cm}}}
\caption[ ]{These diagrams of V--R vs. R--I color show the class models of stars (black) and galaxies (grey) on the left, and stars (black) and quasars (grey) on the right to illustrate their location in color space. The colors plotted are photon flux color indices, which are offset compared to astronomical magnitudes, such that Vega has $V-R= -0.41$ and $R-I= -0.61$. \label{3class}}
\end{figure*}

The shape of the filter transmission curves needs to be known precisely, and is in the best case measured within the imaging instrument itself under conditions identical to the real imaging application. This is easily possible with, e. g., the Calar Alto Faint Object Spectrograph (CAFOS) at the 2.2m telescope on Calar Alto, Spain: in this instrument light from an internal continuum source is sent first through the filterwheel and second through the grism wheel before reaching the detector. Images are taken with and without the filter, so their ratio gives immmediately the transmission curve. Colors measured in narrow filters depend sensitively on the transmission curve, whenever strong spectral features are probed, e.g. the continuum drop at the Ca H/K absorption or the Mg I absorption in late-type stars. In these cases the curve needs to be known rather precisely, since otherwise the calibration would be off, and misclassifications could occur.

\subsection{Potential improvements on the classification}

The quality of the classification reached depends on just the three elements of the method: the quality of the measured data, the choice of the classifier and the quality of the libraries forming the knowledge database for the comparison. In principle, improvements on the performance can be achieved only in the following respects:

\begin{itemize}
\item improvements on the data: the filter set could be changed and potentially be tailored to a specific goal of the survey; the exposure time could be increased or distributed better among the individual filters; the accuracy of the calibration could be increased;

\item improvements on the classifier: the classifier can not be improved fundamentally. A very crucial ingredient for a statistically correct result is a valid assessment of the measurement errors since they form a basic input to the probability calculation. Some simplifications have been introduced which make a difference only among faint objects which are hardly classifiable anyway. For specific goals the classifier can be modified accordingly, and diverse goals potentially ask for contrary strategies. A global maximum in the classification reliability is best achieved by weighting rare classes lower, while searching for rare objects would profit from weighting them higher.

\item improvements on the libraries: this is the most important aspect, since any library-based classifier can obviously recover identifications only if they are contained in the library, and if their spectral profile is well-known. Altogether, this work has strongly benefitted from templates and libraries published in the literature \cite{Fra91,Kin96,Pic98}, which we could arrange into an ordered database. Their limitations are discussed in the respective section. Our present experience suggests that empirical spectra work better for our survey data than purely theoretical spectra. In the future, we would also like to check some of our model assumptions used in the libraries against observations.
\end{itemize}

\section{Simulation of competitive filter sets}\label{simu_comp}

Initially, it should be natural to assume that surveys with different filter sets show quite a different performance in terms of classification and redshift estimation. If a survey aims for  objects with very particular spectra, the filter set can certainly be tailored to this purpose. If the objects of interest span a whole range of spectral characteristics, it is not trivial to guess via analytic thinking which filter set performs best. 

Originally, this method was developed for CADIS using real CADIS data to test it. Then, we intended to optimize it and try to draw conclusions about survey strategies. Aiming for more insight into the question of filter choice, we performed Monte-Carlo simulations on different model surveys by feeding simulated multi-color observations of stars, galaxies and quasars into our algorithm. Here, we present a comparison of three fundamentally different filter sets and show their resulting performance for classification and redshift estimation. 

The three model surveys spend the same total amount of exposure time on different filter sets, but use the same instrument, telescope and observing site. We chose the Wide Field Imager (WFI) at the 2.2-m-MPG/ESO-telescope on La Silla as a testing ground, because it provides a unique, extensive set of filters ranging from several broad bands to a few dozen medium bands to choose from. Furthermore, the WFI is a designated survey instrument which is extensively used by the astronomical community.

\begin{table}
\caption{Filters and 10-$\sigma$-magnitude limits for the three survey setups compared with Monte-Carlo simulations. The I-band filter is a long wavelength passband filter with a cut-on wavelength roughly at 780\,nm. Its far-red sensitivity limit is given by the dropping quantum efficiency of the CCDs. All filters are installed in the Widefield Imager at the 2.2m-MPG/ESO telescope at La Silla observatory. \label{extime}}
\begin{tabular}{lllll}
$\lambda_{cen}$/fwhm (nm) & name & $m_{lim,A}$ & $m_{lim,B}$ & $m_{lim,C}$ \\
\noalign{\smallskip}
\hline
\noalign{\smallskip}
364/38  	& U & 23.5 & 23.5  & 24.1  \\
456/99  	& B & 25.0 & 25.0  & 25.6  \\
540/89  	& V & 24.5 & 24.5  & 25.1  \\
652/162 	& R & 24.5 & 24.5  & 25.1  \\
850/150*	& I & 23.0 & 23.0  & 23.6  \\
\noalign{\smallskip}
420/30	& & 23.6 & 23.98 &   \\
462/14	& & 23.5 &       &   \\
485/31	& & 23.4 & 23.78 &   \\
518/16	& & 23.3 &       &   \\
571/25	& & 23.2 & 23.58 &   \\
604/21	& & 23.1 &       &   \\
646/27	& & 23.0 &       &   \\
696/20	& & 22.8 & 23.18 &   \\
753/18	& & 22.7 &       &   \\
815/20	& & 22.6 & 22.98 &   \\
856/14	& & 22.5 &       &   \\
914/27	& & 22.4 & 22.78 &   \\
\noalign{\smallskip}
\hline
\end{tabular}
\end{table}

\subsection{Filtersets and exposure times}\label{simu_sets}

The three modelled surveys, here called setup ``A'', ``B'' and ``C'', each spend 150\,ksec of exposure time distributed on the following filters (see also Tab.\,\ref{extime}): 

Setup A spends 50\,ksec on the five broad-band filters of the WFI (UBVRI) and 100\,ksec on twelve medium-band filters. Using ESO's exposure time calculator V2.3.1 for the WFI, we related exposure times to limiting magnitudes assuming a seeing of $1\farcs4$, an airmass of 1.2, point source photometry and a night sky illuminated by a moon three days old. The exposure times are distributed such, that a quasar with a power-law continuum $f_\nu = \nu^\alpha$ and a spectral index of $\alpha = -0.6$ is observed with a uniform signal-to-noise ratio in all medium bands. As a result, the twelve medium bands each deliver a 10-$\sigma$ detection of an $R=23.0$-quasar.

\begin{figure*}
\centerline{\hbox{
\psfig{figure=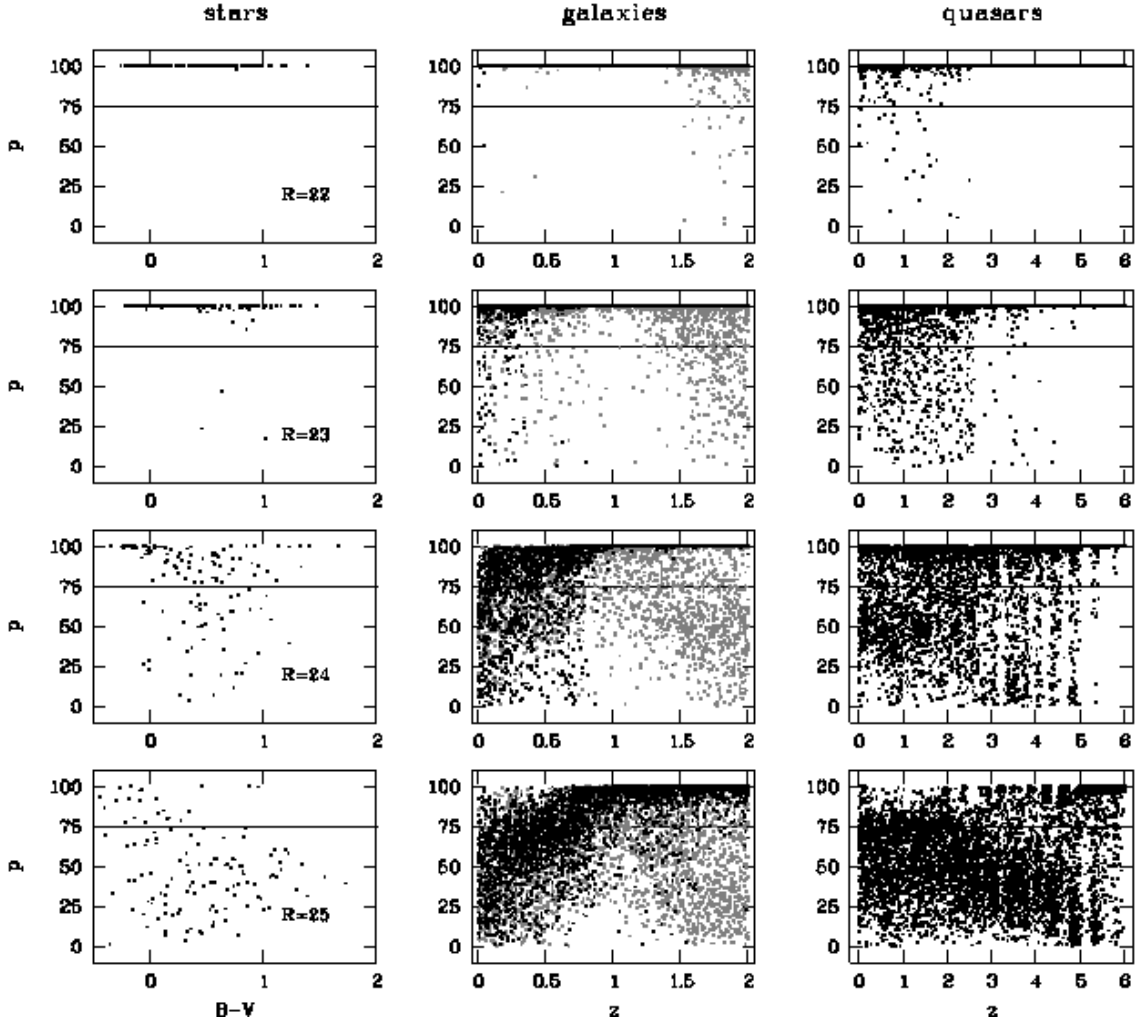,angle=270,clip=t,width=15cm}}}
\caption[ ]{Monte-Carlo simulation for the classification of stars, galaxies and quasars with setup A and $R=22\ldots25$. The probability for a simulated object to be assigned to its original class is plotted over the color $B-V$ for stars and over the redshift for galaxies and quasars, where $B-V$ is an astronomical magnitude. In case of the galaxies black dots denote quiescent galaxies (SED$<$60) and grey dots are starburst systems (SED$\ge$60). For bright objects the performance is limited by a systematic uncertainty of 3\% assumed as a minimum error for the color indices.  \label{clp}}
\end{figure*}

Setup B spends 50\,ksec on the same broad bands but concentrates the 100\,ksec for medium-band work on only six filters reaching a uniform 10-$\sigma$ detection of a $R=23.38$-quasar then. 

Setup C finally spends all 150\,ksec on the broad-band filters and omits the medium bands entirely. 

In Sect.\,\ref{class_a} and Sect.\,\ref{photoz_a} we present the performance results for setup A, which has actually been used for a recent multi-color survey \cite{Wolf00b}. The relative performance of the three setups is compared in Sect.\,\ref{comp_setups}. In Sect.\,\ref{analyt} we attempt to derive some basic analytic conclusions. 

The simulations are carried out by creating a list of test objects from the color libraries presented in Sect.\,3. We assume a certain R-band magnitude and calculate the individual filter fluxes and corresponding errors for each object. Then we scatter the flux values of the objects according to a normal distribution of the flux errors. Finally, we recalculate the resulting color indices and index errors and use this object list as an input to the classification. 

For the stars we use just 131 test objects as there are members in the library. For the test galaxies we take only every third member of the present library giving us 6700 objects. From the quasar library we use every seventh object resulting in 6450 quasars per test run.  

These simulations show us how well the classification can possibly work, assuming that real objects will precisely mimic the library objects. Every real situation will contain differences between SED models and SED reality, sometimes called ``cosmic variance'', which will worsen the performance of every real application. Nevertheless, the simulation highlights the principal shortcomings of the method itself and the chosen filter set in particular. Therefore, it can be used to judge the relative performance of different filter sets. 

We run these tests for stars, galaxies and quasars with magnitudes of $R=$22, 23, 24 and 25, respectively, in order to see how the classification performance degrades from optimum to useless with decreasing object flux. Given that $R=23$ corresponds roughly to the 10-$\sigma$ limit of setup A, the most shallow survey, we expect that the classification has almost reached its best performance at $R=22$. This is due to our assumption of a 3\% uncertainty in the calibration, which causes even the brightest objects with the best photon statistics to perform not much better than an object detected only at a 30-$\sigma$ level. Finally, at $R=25$ objects are well detected only in the broad-band filters, while the medium bands yield only fluxes with errors higher than 40\%. We expect the surveys to be almost useless at this level.

\subsection{Classification performance for setup A}\label{class_a}

\begin{table}
\caption{Classification matrix for objects of $R=23$ in setups A and C as derived from Monte-Carlo simulations. An input vector containing a true number distribution of objects among the three object classes would be mapped by this matrix onto a classified distribution among four classes. Numbers below 0.005 are left blank. \label{cmatrix}}
\begin{tabular}{l|ccc|ccc}
$R=23$	& \multicolumn{3}{c}{true class, setup A} & \multicolumn{3}{c}{true class, setup C} \\
{\it classified as} & star	& galaxy	& quasar & star	& galaxy	& quasar \\
\noalign{\smallskip}
\hline
\noalign{\smallskip}
%		&	&		&	&	&		&	\\
star		& 0.98	& 		&	& 0.96	& 		& 0.03	\\
galaxy		& 0.01	& 0.95		& 0.01	& 0.01	& 0.92		& 0.01	\\
quasar		&	& 0.01		& 0.94	&	& 		& 0.84	\\
unclassified 	& 0.01	& 0.04		& 0.05	& 0.03	& 0.08		& 0.12	\\
\noalign{\smallskip}
\hline
\end{tabular}
\end{table}

We now look at the classification performance as achieved in setup A, the model survey with the highest number of filters, but the shallowest exposures in terms of photon flux detection:

\begin{figure}
\centerline{\hbox{
\psfig{figure=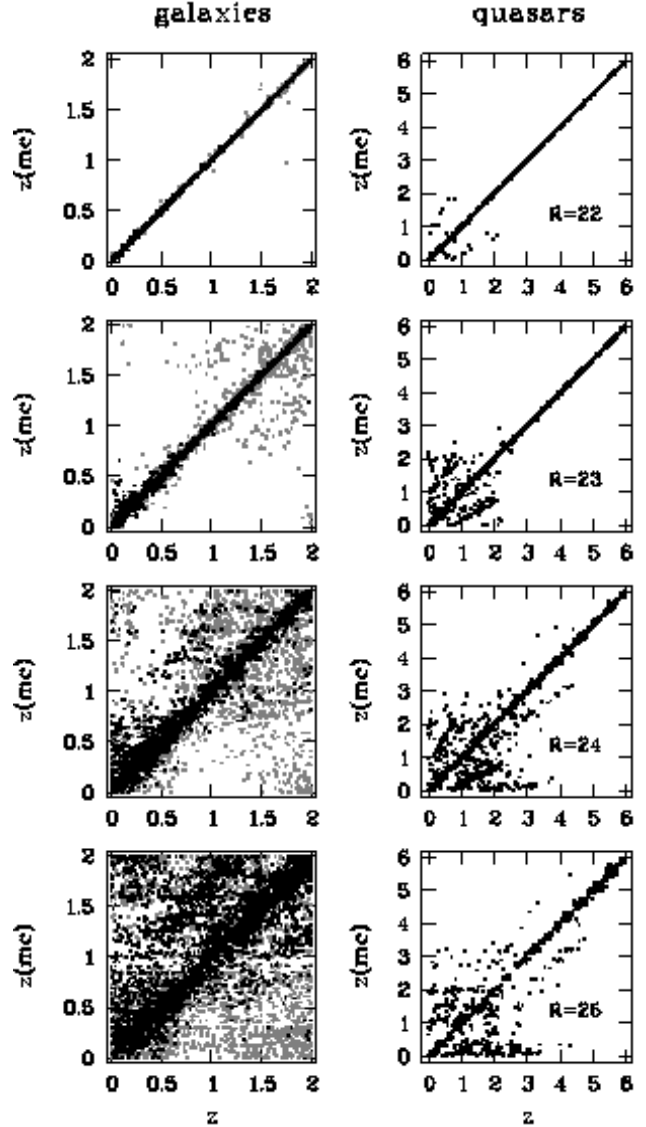,angle=270,clip=t,width=8.3cm}}}
\caption[ ]{Monte-Carlo simulations for the multi-color redshifts of galaxies and quasars with $R=22\ldots25$ according to the Maximum Likelihood estimate in setup A. In case of the galaxies black dots denote quiescent galaxies and grey dots are starburst systems. This diagram shows the redshift estimates for all galaxies, however they were classified, but only for those quasars passing the classification limit of 75\%. \label{zz}}
\end{figure}

For $R=22$ it turned out, that the classification works almost perfect (see uppermost row of diagrams in Fig.\,\ref{clp}). Generally, more than 99\% of all test objects in any class are correctly classified. 

At $R=23$, usually less than 5\% of all objects in any class get lost to unclassifiability. Most affected with 10\% incompleteness are quasars at $z<2.5$ with red spectra and weak emission lines. In this simulation, their location in color space overlaps with starburst galaxies at redshift $1.6<z<2.0$. So far, our galaxy templates contain no information in the spectral range bluewards of the Lyman-$\alpha$ line leaving their U-band flux blank in this redshift range. As a result, the classification omits this band for the comparison with the library galaxies.

At $R=24$, about one third of the stars get lost. These are mostly yellow stars which are too faint in every filter to be classified unambiguously. Rather blue and rather red stars are still successfully classified, because either on the blue or on the far-red side of the filter set they still show significant fluxes and sufficiently accurate color indices. About a quarter of the galaxies would be missed, which are either blue galaxies not showing strong continuum features or red galaxies at redshifts low enough to render them faint in the far-red filters, too. Also, a quarter of the quasars is lost, either red $z<2.5$-quasars overlapping again with starburst galaxies at $1.6<z<2.0$, or $z>2.5$-quasars with weak emission lines overlapping with early-type galaxies at $z<0.4$. 

At $R=25$, the classification has finally become highly incomplete, but can still find very blue stars and very red extragalactic objects like quiescent galaxies and quasars at higher redshift (see bottom row in Fig.\,\ref{clp}, see also Fig.\,\ref{q3} for precise numbers). 

In all simulations, most incorrectly classified objects are {\it unclassifiable} and a minority of them {\bf are} scattered into another class (see also classification matrix, Tab.\,\ref{cmatrix}). Especially, quasars seem to be not strongly contaminated by false candidates. At any magnitude in any setup, less than 1\% of the galaxies are scattered into the quasar candidates except for setup C at $R=25$. Still, this contamination in the quasar class is not negligible, since a minor fraction of a rich class can be a large number in comparison with a poor class. In CADIS we found about 3\% of the extragalactic objects at $R<23$ to be quasars. A contamination of less than 1\% means that less than a quarter of the quasar candidates should be galaxies.

\subsection{Multi-color redshifts in setup A}\label{photoz_a}

Fig.\,\ref{zz} displays the comparison of the photometric MEV+ redshift estimates in setup A with the original true redshifts of the simulated objects. At $R=22$ (see uppermost row of Fig.\,\ref{zz}) the redshifts work quite satisfactorily for galaxies and quasars, which is demonstrated by nearly all objects residing on the diagonal of identity. 

Towards fainter magnitudes, the galaxy redshifts degrade first at both the lower and the higher redshift ends. The deepest working magnitudes are reached in the redshift range of $0.5<z<1$. This feature is due to the location of the 4000\,\AA-break: When the break is located in the central wavelength region of the filter set, many filters are available on either side of the break to constrain its location rather well even for noisy data. For $z=0.15 \ldots 1.15$, the 4000\,\AA-break is at least enclosed by mediumband filters. But if the break is located close to the edge of the filter set and, e.g., detected only by a noisy signal from a single filter, the true redshift interpretation can not be distinguished well from other options. 

Quiescent galaxies still work reasonably fine at the higher redshift end, because they are brighter in the far-red filters. Starburst galaxies mostly degrade at higher redshift, because they have less discriminating (and trustworthily known) features in the UV than in the visual wavelength range. 

The quasar redshifts remain rather precise at $z=2.2 \ldots 6.0$, all the way down to $R=25$. This is the redshift range, where the continuum step over the Lyman-$\alpha$ line can be seen by the filter set and redshift estimates are expected to reach deep. Of course, at $z \ga 4$ the R-band magnitude of quasars appears artificially faint, since it is strongly attenuated by the Lyman-$\alpha$ forest, but the redder filters contain higher flux levels sufficient to constrain the location of the continuum step. Redshift confusion arises first in the low-redshift region working its way up to higher redshifts with decreasing brightness. At $z<2.2$ the continuum shows no Lyman-$\alpha$ forest in our filter sets, but only a redshift invariant power-law shape. In this case, the multi-color redshifts rely solely on some emission-lines showing up in the medium bands.

\begin{figure}
\centerline{\hbox{
\psfig{figure=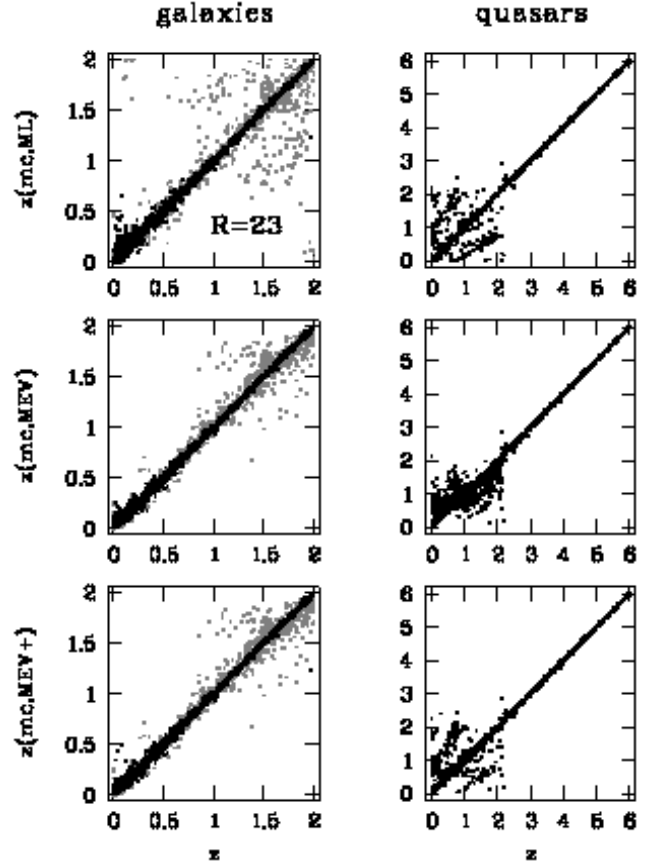,angle=270,clip=t,width=8.3cm}}}
\caption[ ]{Monte-Carlo simulations for multi-color redshifts of galaxies and quasars with setup A and $R=23$ according to the estimators Maximum Likelihood (ML), Minimum Error Variance (MEV) and our advanced MEV with better handling of bimodalities (MEV+). In case of the galaxies black dots denote quiescent galaxies and grey dots are starburst systems. Shown are all galaxies, but quasars only if they passed the classification limit of 75\%. It seems that ML and MEV+ are almost equivalent for quasars, while for galaxies MEV and MEV+ make no visible difference. Objects considered uncertain by the MEV estimator do not get an MEV estimate assigned, but they receive an ML estimate that can potentially be wrong. \label{z3}}
\end{figure}

\begin{figure}
\centerline{\hbox{
\psfig{figure=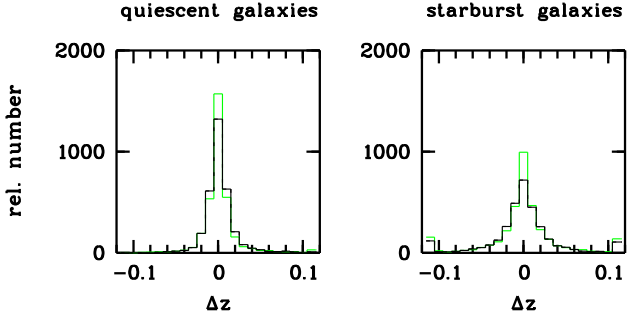,angle=270,clip=t,width=8.3cm}}}
\caption[ ]{Distribution of true redshift estimation error ($\Delta z = z_{mc}-z$) among simulated galaxies with $R=23$ in setup A separated for quiescent and starburst objects. The solid line shows results for the Minimum Error Variance estimators (MEV and MEV+ are virtually the same) and the grey line those for the Maximum Likelihood estimator (ML). Starburst systems show higher errors and some large mistakes with $\Delta z>0.1$. \label{zh3}}
\end{figure}

Some concentrated linear structures are visible off the diagonal at lower redshift with the best contrast at $R=24$. Their origin is a misidentification of weak emission lines: There are two structures mirrored at the diagonal following the linear relations $(1+z_{phot})/(1+z) \approx 1.74$ and $(1+z)/(1+z_{phot}) \approx 1.74$. They are caused by a confusion of the Mg\,II line with the H$\beta$ line. Another structure at $(1+z_{phot})/(1+z) \approx 1.25$ arises from weak Lyman-$\alpha$ lines of very blue quasars which are interpreted as C\,IV lines, or weak C\,IV lines which are taken to be C\,III lines. The extent of these structures across the diagram obviously depends on the visibility of the involved lines within the medium-band filter set. Finally, there is a large group of quasars estimated to be at nearly zero redshift, but truely strechting even beyond $z=3$. These are among the quasars with the lowest emission line intensities in the library, which basically display only their redshift-invariant power-law continuum in the filters.

\subsection{Maximum Likelihood redshift versus Minimum Error Variance redshift}\label{mlvsmev}

We now compare the relative performance of three different redshift estimators using the example of galaxies and quasars at a fixed magnitude of $R=23$. We have used the Maximum Likelihood (ML) method, the Minimum Error Variance (MEV) method and an advanced MEV method (MEV+) as we defined it in Sect.\,\ref{algor_applic}. 

While the Maximum Likelihood (ML) method always gives a redshift estimate, the Minimum Error Variance (MEV) method does not in the way we use it. Some objects have probability distributions which are close to flat yielding a redshift estimate that reflects primarily the redshift interval chosen for the template library rather than giving a reliable physical interpretation of the object. We do not assign any estimate to these {\it uncertain} objects (as we defined them in Sect.\,2.4), which is justified with their estimates being senseless anyway. A caveat for a direct performance comparison is the fact, that the MEV/MEV+ methods ignore the {\it uncertain} objects, whose selection function is redshift-dependent at the faint end and could furthermore be different in a real dataset due to cosmic variance.

As shown in Fig.\,\ref{z3}, the different estimators deliver rather comparable results with quite similar redshift accuracy. In case of the quasars the improved MEV+ method (which can detect bimodal probability distributions) performs different from the standard MEV method but rather similar to the ML method. This is due to bimodalities where the MEV estimate is a weighted average of the two present probability peaks, while the MEV+ estimate decides for the single peak containing the higher probability integral, which is likely to be roughly coincident with the ML estimate pointing at the redshift with the highest individual probability. Bimodalities can again be seen as linear structures off the main diagonal and arise from confusion among emission lines. In case of the pure MEV method the peak associated with the wrong solution is averaged with the correct solution residing on the diagonal, and the MEV plot shows smeared out structures around the diagonal rather than the linear ones like the ML or MEV+ plots.

\subsection{The three setups in comparison}\label{comp_setups}

\begin{figure*}
\centerline{\hbox{
\psfig{figure=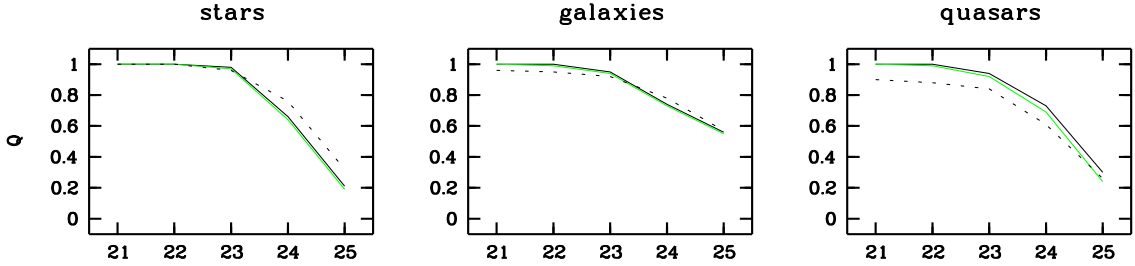,angle=270,clip=t,width=15cm}}}
\caption[ ]{Fraction Q of simulated objects which are correctly classified in the three different setups (solid line = setup A, grey line = setup B, dashed line = setup C). Except for faint stars, setup A and B are most successful. \label{q3}}
\end{figure*}

\begin{figure*}
\centerline{\hbox{
\psfig{figure=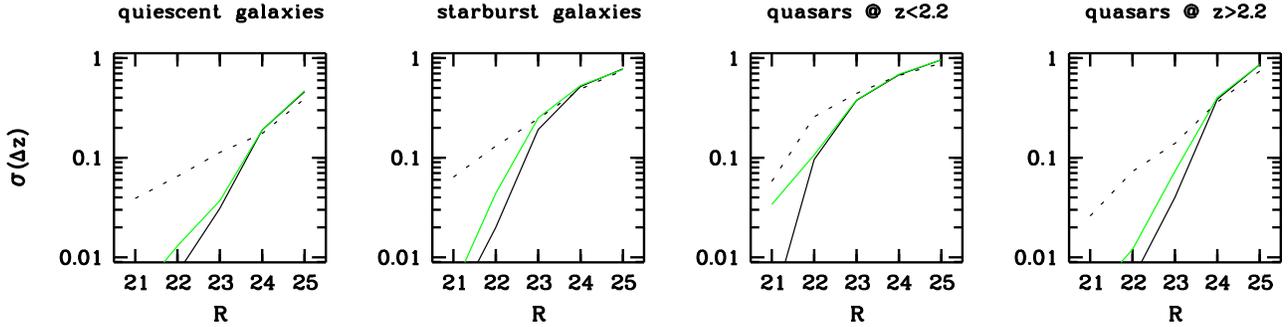,angle=270,clip=t,width=17cm}}}
\caption[ ]{Variance of true redshift estimation error ($\Delta z = z_{mc}-z$) among simulated objects in the three different setups (solid line = setup A, grey line = setup B, dashed line = setup C) based on Maximum Likelihood estimate. Setups A and B provide the highest redshift resolution. Early type galaxies work better due to their higher continuum contrast at the 4000\,\AA-break. Nearby quasars without continuum features do not work too well, since the redshift estimate has to rely on emission lines. \label{sz22}}
\end{figure*}

All setups are designed to spend the same amount of exposure time on a survey field, but distribute it on different filter sets. The pure broad-band survey, setup C, collects far more photons than the setups A and B, which are mainly exposing medium-band filters. But due to their higher spectral resolution, we expect setups A and B to contain more information per photon.

In fact, it turns out, that the classification performance of all three setups is quite similar, which implies that the lack of photons in the medium bands is pretty much compensated by their higher information content (see Fig.\,\ref{q3}). Among the small remaining differences, there is a tendency for the medium-band setups to be more efficient in finding quasars, supposedly because their spectra contain emission lines which are more prominent in narrow filters. 

Also, there is a slight trend indicating that the medium-band surveys sustain a high level of completeness to somewhat fainter magnitudes and then drop more sharply than setup C. In the incompleteness range of very faint magnitudes, all setups perform rather equally meager.

The same trends are more clearly present among the multi-color redshifts, where we compare the statistics for the ML estimator (see Fig.\,\ref{sz22}): Setups A and B provide a much better redshift resolution at the usual working magnitudes. They only fall behind the performance of setup C by a rather insignificant degree in the faintest regime, where the redshift estimates are close to unusable to start out with. This advantage of setup C results just from the broad bands being deeper by $0\fm6$, where the medium-band filters do not contribute to the result anymore.

For brighter objects, estimates in setup A are better than in setup B by an average factor of two, just reflecting the difference in spectral resolution. After all, the convolution of any measurement with a 0\fm03-Gaussian (to account for the calibration errors) makes better photon statistics useless among objects, which are detected at more than a $\sim 30\sigma$-level. Thus, only increasing the number of filters improves the result for these objects while increasing the depth of any filter has no effect.

At this point we like to emphasize, that the calibration uncertainty limits the best achievable performance. We stress, that a large calibration error of e.g. 10\% would turn an entire survey catalog into a collection of ``less-than-10-$\sigma$-objects'', at least within our method. If calibration is expected to be a problem due to instrumentation or observing strategy, this conclusion strongly suggests that a large number of filters giving many noisy datapoints deliver more information than a few long exposed and formally deep filters that can not exactly be matched together.

Once more we look into the details within classes: It is no surprise that quiescent galaxies with rather prominent 4000\,\AA-breaks receive more accurate redshift estimates than starburst galaxies with less contrasty continuum features. When comparing equal accuracies, we find that estimates for quiescent galaxies reach typically one magnitude deeper than for starburst objects. When aiming for a redshift resolution of $\sigma_z \approx 0.03$ among quiescent galaxies, it is interesting to see, that any of the medium-band surveys reaches two magnitudes deeper than the broad-band survey (setup C).

The quasar redshifts work best at $z>2.2$, when the estimation depends not only on emission lines but can take advantage of a strong continuum feature being present within the range of the filter set, i.e. the continuum suppression bluewards of the Lyman-$\alpha$ line. As in the case of galaxies, setups A and B have significantly stronger resolving power in terms of redshift than setup C, with setup A again being the best choice. 

It is inspiring to conclude from these simulations, that photometric redshifts for quasars are feasible and are supposed to reach accuracies of $\sigma_z \la 0.1$ in surveys with medium-band filters. Furthermore, observations from the CADIS survey find a surprising number of faint quasars, whose multi-color redshifts were indeed proven by spectroscopy to be as accurate as expected from the simulations (see Wolf et al. 1999 and paper II). 

Altogether, setup A seems to be the most successful among the ones discussed for photometric classification and redshift estimation. It has no disadvantages compared to the other setups, especially it does not lack working depth compared to the pure broad-band survey. Viewing the almost vanishing differences between setup A and B, there might be no incentive to increase the number of filters even higher.

Still, setup A shows a selection function for a successful classification with some redshift dependence. Among the quasars shown in Fig.\,\ref{clp}, we can see some vertical stripes containing objects at selected redshifts, which are not successfully classified anymore, while the neighboring redshifts still work well. In principal, a set of neighboring medium-band filters touching in wavelength and covering the important spectral range completely would most likely result in a selection function with the smoothest shape and smallest redshift dependence.

\subsection{Analytic thoughts on filter choice}\label{analyt}

In this section, we would like to address the issue of choosing an optimal filter set by analytic thoughts based on a simplified picture of the classification problem. We assume, that we are still limited by a fixed amount of telescope time, which we can distribute over some filters. If different wavebands could be imaged simultaneously, it would be obvious that even a faintly exposed full-resolution spectrum would be better than an unfiltered white light exposure, as long as read-out noise of the recording detector is not an important constraint. Here, we want to discuss the less obvious scenario of consecutive exposures in different wavebands. 

As mentioned in the introduction, the choice of the optimum filter set depends entirely on the goal of the survey. For surveys aiming at a particular type of objects with characteristic colors, tailored filter sets can be designed. But if we intend to integrate different survey applications into one observational program on a common patch of sky, then we need a single survey to identify virtually every object unambigously. In this scenario two choices have to be made:
\begin{enumerate}
\item Assuming constant filter width, the choice between \\
         a) either fewer filters with more photons each \\
         b) or more filters with fewer photons each.
\item Assuming constant filter number, the choice between \\
         a) broad filters with more photons and less resolution \\
         b) narrow filters with less photons and more resolution.
\end{enumerate}

We first note, that if all colors were equally discriminating for each object, the choice would be arbitrary. Any distribution on any number of equally wide filters would provide the same total discriminative power and classification performance. In practice, objects can reside at many different redshifts and usually only part of their spectra have discriminating features. 

% But if we intend to integrate different survey applications into one observational program on a common patch of sky, then we need a single survey to identify virtually every object by using a filter set which enables understanding the different objects above some magnitude limit. In this case, the flux detection limit of individual bands are not the key numbers any more, but instead the limit of a successful classification as needed for various science applications. If the classification takes all available color data into account, like template fitting procedures do, then the flux limit of a single filter is not the only relevant number, since the performance will depend to a large extent on the filter choice.

% We do not want to confuse the power of a survey to perform a template-based classification with its flux detection limits! Especially, we should not get distracted by the deeper flux limits of single filters in a broad-band survey compared to a medium-band survey. Asking for this number in a set with five filters means asking for 20\% of the information obtained in the survey, while in a set of 20 filters the respective flux limit tells about the information content of only 5\% of the survey. 

We now try to obtain some insight into this question based on very basic template assumptions. For simplicity, we now just assume two different possible objects posed to the classification algorithm, with one of them being a quasar only distinguished by an emission line from another object with an otherwise identical spectrum. 

Addressing choice (1), we find, that concentrating on few filters would mean that only few quasars display their emission line in a filter and can be classified correctly down to some limit, while many objects would be unclassifiable. The classification would lack completeness, but reach deep for a few objects. Distributing exposure time among many filters covering the entire spectrum would give every quasar a chance to show its emission line, which implies that every object is well classifiable but not to the same depth. The classification would remain rather complete and degenerate more sharply than in the case of few filters when reaching its limiting magnitude. 

Addressing question (2), we assume one of the filters to observe the emission line and evaluate the line contrast obtained. As long as the line is completely contained in the filter bandpass, our signal, i.e. the absolute flux difference to the continuum induced by the line, is a constant value irrespective of the filter width. The noise is given by the square-root of the total flux from the object which increases along with the width of the bandpass. The optimum signal-to-noise ratio is obtained with a filter matching just the width of the emission line. Any narrower filter would cut off line flux, thus shrinking the signal more than the noise. 

Using both conclusions we can ask for the optimum strategy when aiming for high sensitivity and completeness across some redshift range. This goal requires that we observe the emission line in any case regardless of the redshift. Therefore, we need $n$ filters to cover the entire spectrum in question, depending on the filter width $\Delta \lambda \propto 1/n$. Given a fixed total amount of exposure time, the exposure time per filter and thereby the counts measured from the line are $S_{line} \propto 1/n$. The total flux $S_{tot}$ in this filter depends on the same exposure factor and on $\Delta \lambda$, so that the $S_{tot} \propto 1/n^2$ and $\sigma_{tot} \propto 1/n$. Therefore, the signal-to-noise ratio $S_{line}/\sigma_{tot} =$ const, independent of the number of filters in any set providing complete coverage. 

In summary for the simple quasar example, we have a free choice on the filter set, as long as we cover the spectrum. It seems, that the width of the filters does not affect the magnitude limit for a successful classification, but it determines directly the redshift resolution. Having the free choice, many filters tailored to the typical width of quasar emission lines would be the best solution.

Another example is photometric star-galaxy separation. Some red stars display broad-band colors similar to some redshifted early-type galaxies. Good photometric accuracy is required to tell them apart, especially if only few filters are available. With medium-band filters enclosing the redshifted 4000-\AA-break of the galaxies and probing the absorption bands of stars, the two classes can easily be discriminated even at rather noisy flux levels.

Let us assume the most general imaginable case for the classification problem, where the object spectra can have features with potentially any location and any width (due to redshift as well as class). The arbitrary location calls for a filter set covering the entire spectrum. Again, we are left with the choice of many narrow versus few broad filters mentioned in the simple quasar example just above. And again, as long as the features are smaller than the filter width, the choice of filters makes no difference to the classification, if the same total amount of telescope time is used.

We now consider an abstract information value $I$ obtained by a survey. It depends on the number of filters $n$, on the photons collected in each of them $N_{ph} (f)$ and on the information $I/N_{ph} (f)$ that a single photon carries after passing through a given filter. If on average the same amount of information is obtained in every filter, we get:
\begin{eqnarray}
	I = n \times N_{ph} (f) \times \frac{I}{N_{ph}}(f) ~.
\end{eqnarray}

For complete coverage the number $n$ of filters again depends on the filter width $\Delta \lambda \propto 1/n$. Given a fixed total amount of telescope time, the exposure time per filter is $\Delta t \propto 1/n$ and thus the number of photons collected is $N_{ph} (f) \propto 1/n^2$. Since narrow filters show features with more contrast than broad filters, we can assume that the information per photon is inversely proportional to the filter width: $I/N_{ph} (f) \propto 1/\Delta\lambda$, and thus $I/N_{ph} (f) \propto n$. Altogether, the information content of the survey results to:
\begin{eqnarray}
	I = n \times 1/n^2 \times n = \rm{const} ~.
\end{eqnarray}

In {\it theory}, the amount of information in terms of classifiability of objects depends only on the total telescope time and not on the characteristic width of the filters, as long as they cover the entire spectral range in question. The smaller number of photons in the medium-band survey is compensated by the larger number of filters and the higher information content per photon. But this conclusion is based on three simplified assumptions:
\begin{itemize}
  \item a much simplified picture of object spectra 
  \item photon noise be the only source of measurement error
  \item libraries resembling true nature accurately and completely
\end{itemize}

In {\it practice}, there are several advantages for medium-band and mixed surveys compared to broad-band surveys, especially when combined with our classification scheme:
\begin{itemize}
  \item medium-band surveys always provide better redshift resolution
  \item medium-band surveys perform much better at the limit of calibration accuracy
        by providing many more datapoints with higher spectral resolution 
  \item medium-band surveys can tolerate inaccurate libraries and {\it cosmic variance}
        much better for the same reason
  \item medium-band surveys sampling the the wavelength range somewhat sparsely (as the model
        surveys A and B in our simulations) can have their filters placed to avoid strong
        night sky emission lines and to suppress background noise
\end{itemize}

Especially the last three advantages can cause a medium-band survey to reach even deeper than a broad-band survey in terms of classification and redshift estimation, although its nominal flux detection limits might have suggested inferior performance to the intuitive judgement. 

The disadvantage of a survey project involving many medium-band filters is, that it needs a larger minimum amount of telescope time, since a few constraints in observational strategy have to be met. An optimal survey has requirements for:
\begin{itemize}
  \item a minimum number of exposures per filter to eliminate a fringe pattern
        and to close gaps in a CCD mosaic
  \item a minimum exposure time for every frame in order to avoid being 
        limited by read-out noise 
  \item a minimum number of filters derived from a certain minimum coverage
        of a wide spectral range.
\end{itemize}

\section{Conclusions for real multi-color applications}

\subsection{Calibration of colors}\label{calib_data}

Obviously, the measurements also need a careful calibration among the wavebands. A large erroneus offset can be disastrous for the photometric classification of narrow class structures in color space. If, e.g., true stars were measured with shifted colors, the classification would potentially find it rather in the location of library galaxies or quasars, and vice versa. Also, the redshift estimates would be thrown off by color offsets.

Calibration problems are of greatest concern, when rare objects are searched and their class gets contaminated. Especially, when class volumes are almost touching in color space, already small calibration errors can push objects into the wrong class. E.g., in many filter sets the quasar class is not well separated from stars and galaxies. In the presence of a calibration error, abundant galaxies can be pushed into the quasar class potentially making up the largest population among the precious candidates. The shape of class volumes is likely to cause quite some redshift dependence for the contamination. Then objects in some redshift range can become virtually unidentifiable, if they are overwhelmed by contaminants.

If calibration errors were known and quantified, they could as well be removed. If they were present but not realized, the measurements would look too accurate and a seemingly faithful classification would be derived, which is potentially wrong. Thus, as long as the calibration errors are unknown, it is still important to take their potential size into account for the error estimates on which the classification is based. As a result, the performance of the classification for bright objects is indeed limited by the calibration error. 
 
We assume calibration errors on the order of 3\% for the colors in our surveys, which implies that the quality of the classification saturates for objects that are more than $1\fm5$ brighter than the 10-$\sigma$-limits of the survey. On the other hand, if we assume for the moment poor data reduction or uncorrected galactic reddening changing the colors by, e.g., 10\%, this would turn an entire survey catalog into a collection of ``less-than-10-$\sigma$-objects'' --- a devastating effect for the survey quality. 

An accurate relative calibration among many wavebands is best ensured by establishing a few spectrophotometric standard stars in each of the survey fields, a successful approach that we have made into a standard procedure in CADIS. This task can be carried out in a photometric night by taking spectra of the new standards and connecting these to published standard stars \cite{Oke90}. This way, spectrophotometric standards are available in every one of the survey exposures, which will not require any further calibartion efforts regardless of the conditions under which the regular imaging is carried out. Obviously, standard star spectra are supposed to cover the entire filter set, but if a mixture of (e.g. optical and infrared) instruments is used, the calibration will involve different procedures to be matched.

\subsection{The optimum survey strategy}

The most basic result of our study on the performance of different multi-color surveys is, that even for small systematic errors in the color indices of $s_{g-h} = 0\fm03$, a survey with 17 bands performs better in classification and redshift estimation than one with only few bands. For the 17 band case we found that the limiting magnitude for {\it reasonable} performance is reached when the typical statistical (i.e. photon noise) errors are on the order of 10\%. It is obvious, that larger systematic errors will worsen the performance and will allow even higher statistical errors before the survey deteriorates significantly. For the survey strategy this implies that pushing the statistical errors in each band well below the systematic errors will add nothing to the survey performance. When $\Delta t_{int}$ is the integration time required to reach $\sigma \approx 1/2 s_{g-h}$, the optimum number of bands $N$ for a given amount of total time $t_{tot}$ is roughly 
\begin{eqnarray}
	N = t_{tot} / \Delta t_{int} ~. 
\end{eqnarray}

Although our present study has been confined to the wavelengths region attainable by optical CCDs and did not address the total wavelength coverage of the survey explicitely, it is predictable that further bands extending the wavenlength coverage (e.g. by adding NIR bands) will have a larger effect than splitting the optical bands. In particular, the maximum redshift for a reliable galaxy classification will be extended.

As the color indices are the prime observables entering the classification and redshift estimation process, it is clear that any multi-color survey has to be processed such, that these indices are measured in an optimum way. For ground-based observations it is of great importance to avoid that variable observing conditions introduce systematic offsets between bands when the observations are taken sequentially. First of all, this requires to assess the seeing point spread function on every dataset very carefully. Second, one has to correct for the effect of variable seeing which might influence the flux measurement of star-like and extended objects in a different way.

In CADIS, we essentially convolve each image to a common {\it effective} point spread function and measure the central surface brightness of each object (see paper II for details). This has the disadvantage, that the spatial resolution (i.e. the minimum separation of objects neighboring each other) is limited by the data with the worst seeing. However, it is not clear whether the obvious alternative --- deconvolution techniques --- can be optimized such that the systematic errors can be kept below a few percent for a wide variety of objects.

The performance of the MEV estimator depends critically on the assumption that not only the color indices but also their errors are determined correctly. For the survey strategy this implies, that an optimization of the photon noise errors under the expense that an accurate estimation of these errors is no longer possible may lead to worse performance than slightly larger errors which are known accurately.

\subsection{Ongoing applications and their scientific goals}

In this section, we want to mention examples for survey applications using this method and comment on the usefulness of our classification approach. A number of multi-color surveys have been conducted, where filters and exposure times were chosen to match some primary survey strategy. Although, none of these might have been optimal choices in terms of a general classification, we used or intend to use our approach to extract class and redshift data on the objects contained. These surveys are in chronological order of their beginning:

\begin{enumerate}
\item The {\it Calar Alto Deep Imaging Survey} (CADIS): Three broad-band and twelve medium-band filters have mostly been chosen to match the needs of the emission line survey in CADIS, while some of them fill in gaps in the spectral coverage. The multi-color part of CADIS has been used to study the evolution of the galaxy luminosity function at $z \la 1$, to search for quasars at all visible redshifts and to use the observed faint stellar population to check models of the Galactic structure and the stellar luminosity function \cite{Mei98,Wolf99,Fri00,Phl00,Wolf00a}.

\item A lensing study of the galaxy cluster Abell 1689: Two broad-band and seven medium-band filters have been chosen to separate well between the cluster galaxies at $z \approx 0.19$ and the background population. The galaxy luminosity function in the background of the cluster is compared to a control field taken from CADIS, and the cluster mass is estimated from weak lensing effects on the apparent luminosities \cite{Dye00}.

\item A widefield project for {\it Measuring Agn redshifts by Medium-Band Observations} (MAMBO): The filters (setup A from Sect.\,\ref{simu_sets}) are chosen to provide a selection function and a redshift accuracy for quasars and galaxies, which is as independent of redshift as feasible. The data will be used to study the faint end of the quasar luminosity function at all accessible redshifts $z \ga 1$ and galaxy-quasar correlation at $z \la 1$, as well as weak lensing effects in the cluster group Abell 901/2 and in the open galaxy field \cite{Wolf00b}.

\item The {\it Sloan Digital Sky Survey} (SDSS): Five broad filters have been chosen, which span the entire range of presently available CCD sensitivity. We intend to apply our classification to search for quasars and to separate stars from compact galaxies, where morphology data are not sufficient.
\end{enumerate}

From simulations of the classification scheme presented in this paper, we expect in all these projects, that we should be able to classify virtually all objects above some magnitude limit purely by color, and that especially the medium-band surveys should have selection functions which are not very dependent of redshift. This way, we can omit morphological criteria for defining catalogs of the stellar vs. galaxy vs. quasar population. This conclusion leads to a number of advantages for our method, we like to state explicitely here:

\begin{itemize}
\item The star-galaxy separation reaches deeper and avoids confusion better than if based on morphology. Accordingly, studies of the stellar population and of the galaxy population can be extended to much fainter magnitude.  

\item Quasars can be found at all accessible redshifts, especially in the medium-band surveys, where the overlap with the stellar sequence at redshifts of $2.2 \la z \la 3.5$ is much reduced in the color space.

\item Quasars do not need to be selected from a subcatalog of point sources only, allowing for objects with resolved host galaxies to be found, including high-redshift Seyfert I galaxies.

\item We also believe that the multi-color redshifts from the medium-band surveys are accurate enough so that the applications listed above do not require follow-up spectroscopy. E.g., the evolution of the galaxy luminosity function can be analysed with the multi-color redshifts, as long as no trends in a redshift intervall $\Delta z < 0.1$ are searched for. 
\end{itemize}

\section{Summary}

We presented an innovative method that performs a multi-color classification and redshift estimation of astronomical objects in a unifying approach. The method is essentially based on templates and evaluates the statistical consistency of a given measurement with a database of spectral knowledge, serving as a second, very crucial input to the algorithm.

The introduction of this method is motivated by the quest for a statistically correct extraction of the information present in the color vectors of surveys with many filters. The method is derived from basic statistical principles and calculates probability density functions for each survey object telling us two different results simultaneously: the class membership and redshift estimates according to the Maximum Likelihood (ML) and Minimum Error Variance (MEV) estimators. We add our own version of the MEV technique featuring improved handling of bimodalities in the probability function.

Our choice for the database is a large, systematically ordered library containing templates for stars, galaxies and quasars, which are supposed to cover virtually all but some unusual members among each of the three object classes. The libraries were established from a few model assumptions and templates published by various authors and extracted from the literature.

The method can be implemented in a computationally very efficient way, by using directly color indices as object features. We showed that our color-based approach is expected to deliver results consistent with those from flux-based template-fitting algorithms. 

The accuracy of the data calibration is a very important issue, constraining the design of the libraries and limiting the maximum achievable performance of the method via the effective photometric quality. Calibration errors can distort results and shrink the information output.

We carried out Monte-Carlo simulations for three model surveys using the same total exposure time but different filter sets. One of them is a UBVRI broad-band survey, while the other two expose two third of the time in various medium-band filters. Altogether, the performance of all three setups was rather similar despite the quite different numbers of collected photons. So it appears, that medium-band filters obtain more information per photon and thereby compensate the loss of depth in terms of flux detection, from which they suffer in comparison to broad bands. Among the differences, medium-band surveys performed better than the broad-band survey for finding quasars, and they provided much higher redshift resolution in their estimates. Also, in the presence of calibration errors or uncorrected reddening effects, bright objects are not easier to classify than faint ones, and a large number of shallow filters might provide more information than a small number of deeply exposed filters.

Based on simple analytic assumptions, we have discussed the relative information content of surveys with different characteristic filter width. All surveys using the same amount of total telescope time and filter sets stretching over the entire spectral range of interest, should perform equal in terms of classification. This theoretical conclusion depends on perfect calibration and perfect template knowledge. 

In practice, the classification should reach deeper in medium-band surveys than in broad-band surveys, because the former are less affected by inaccuracies in the calibration and in the template library. Furthermore, the filters can be chosen to avoid noise from strong night sky emission lines which is not possible with broad-band filters. 

In particular, using the proposed statistical classification approach in a suitable medium-band survey it should be possible

\begin{itemize}
\item to separate stars from apparently compact galaxies down to rather deep limits exceeding the potential of morphological classification,

\item to find quasars rather efficiently and completely, i.e. with very little contamination, and

\item to obtain quite good multi-color redshift estimates with errors on the order of $\sigma_z \approx 0.1$ for $z>2$-quasars, and on the order of $\sigma_z \approx 0.03$ for galaxies. 
\end{itemize}

This method should be very suitable for many survey-type applications, which usually require only low spectral resolution and finite accuracy in the derivation of physical parameters, but aim for large samples to feed statistical studies and to search for rare and unusual objects. Of course, if you need a 100\% sure confirmation on the nature of an individual object, or if you aim for high resolution studies, it gives you only a preselection of candidates.

In paper II we show, that this method is very powerful and indeed of great practical relevance for multi-color surveys with many filters like in the case of CADIS. The results of our shown simulations compare well with the performance of a real survey, and therefore, they can in fact be used for testing the performance of future survey designs. 

{\bf
\begin{acknowledgements}
The authors thank H.H. Hippelein for helpful discussions on template fits and H.-M. Adorf for some on classification methods and their fine-tuning. We also thank D. Calzetti for kindly making available the galaxy templates in digital form. We would finally like to thank the referee, Dr. S. C. Odewahn, for detailed comments improving the paper. This work was supported by the DFG (Sonderforschungsbereich 439).
\end{acknowledgements}
}

\end{document}